\documentclass[sigconf]{acmart}

\usepackage{multirow}
\usepackage{xcolor}
\usepackage{longtable}
\usepackage{graphicx}
\usepackage{lscape}
\usepackage{rotating}

\AtBeginDocument{%
  \providecommand\BibTeX{{%
    \normalfont B\kern-0.5em{\scshape i\kern-0.25em b}\kern-0.8em\TeX}}}

\copyrightyear{2024}
\acmYear{2024}
\setcopyright{acmlicensed}\acmConference[CHI '24]{Proceedings of the CHI Conference on Human Factors in Computing Systems}{May 11--16, 2024}{Honolulu, HI, USA}
\acmBooktitle{Proceedings of the CHI Conference on Human Factors in Computing Systems (CHI '24), May 11--16, 2024, Honolulu, HI, USA}
\acmDOI{10.1145/3613904.3642574}
\acmISBN{979-8-4007-0330-0/24/05}

\newcommand\edit[1]{\textcolor{black}{#1}}
\newcommand\finaledit[1]{\textcolor{black}{#1}}

\begin{document}

\title[AI in Peer Support]{The Role of AI in Peer Support for Young People: A Study of Preferences for Human- and AI-Generated Responses}


\author{Jordyn Young}
\affiliation{%
  \institution{University of Michigan}
  \city{Ann Arbor}
  \state{MI}
  \country{USA}
}
\email{jordyny@umich.edu}
\orcid{0000-0002-1565-0642}

\author{Laala M Jawara}
\affiliation{%
  \institution{Drexel University}
  \city{Philadelphia}
  \state{PA}
  \country{USA}
}
\email{lj466@drexel.edu}
\orcid{0009-0009-2382-7565}

\author{Diep N Nguyen}
\affiliation{%
  \institution{Drexel University}
  \city{Philadelphia}
  \state{PA}
  \country{USA}
}
\email{kn569@drexel.edu}
\orcid{0000-0002-4671-4098}

\author{Brian Daly}
\affiliation{%
  \institution{Drexel University}
  \city{Philadelphia}
  \state{PA}
  \country{USA}
}
\email{brian.daly@drexel.edu}
\orcid{0000-0001-5828-7313}

\author{Jina Huh-Yoo}
\affiliation{%
  \institution{Drexel University}
  \city{Philadelphia}
  \state{PA}
  \country{USA}
}
\email{jh3767@drexel.edu}
\orcid{0000-0001-5811-9256}

\author{Afsaneh Razi}
\affiliation{%
  \institution{Drexel University}
  \city{Philadelphia}
  \state{PA}
  \country{USA}
}
\email{afsaneh.razi@drexel.edu}
\orcid{0000-0001-5829-8004}

\renewcommand{\shortauthors}{Young, et al.}

\begin{abstract}

Generative \finaledit{A}rtificial \finaledit{I}ntelligence (AI) is integrated into everyday technology, including news, education, and social media. AI has further pervaded private conversations as conversational partners, auto-completion, and response suggestions. As social media becomes young people's main method of peer support exchange, we need to understand when and how AI can facilitate and assist in such exchanges in a beneficial, safe, and socially appropriate way. We asked 622 young people to complete an online survey and evaluate blinded human- and AI-generated responses to help-seeking messages. We found that participants preferred the AI-generated response to situations about relationships, self-expression, and physical health. However, when addressing a sensitive topic, like suicidal thoughts, young people preferred the human response. We also discuss the role of training in online peer support exchange and its implications for supporting young people's well-being. Disclaimer: This paper includes sensitive topics, including suicide ideation. Reader discretion is advised.

\end{abstract}



\maketitle

\section{Introduction}

\edit{Today's generation of youth is the first to grow up with ubiquitous internet access. Most} (92\%) young people access the internet daily, with 24\% describing their online presence as "almost constant" \cite{vogels2022teens}. \edit{Support is an important reason why young people engage online \cite{bae2021social, pretorius2019young}. Research suggests this is because online help-seeking interactions offer many benefits, including anonymity and privacy \cite{pretorius2019young, best2016seeking, bradford2014adolescent}, ease of access and immediacy \cite{bell2018suicide, best2016seeking, burns2016role, burns2010internet}, connection with others with similar experiences \cite{frost2015self, horgan2010young, mar2014exploring},
increased perceived control of one's help-seeking journey \cite{pretorius2019young, best2016seeking, mar2014exploring}, capabilities for self-reliance \cite{pretorius2019young, ellis2013young, mar2014exploring}, early access \cite{pretorius2019young, birnbaum2017role, frost2015self}, and inclusivity \cite{pretorius2019young, best2016seeking}. The wide reach of the internet also means that young people can connect with support-givers from various backgrounds. These interactions can be with professionals in specific areas (e.g., medical doctors \cite{khandpur2017telemedicine}), trained support specialists (e.g., people recovered from substance abuse \cite{fruitman2023online}), or peers they may otherwise encounter in school or other parts of daily life \cite{huh2023help}.} 

\edit{Now, a new evolution of technology is entering the landscape of online support -- \finaledit{A}rtificial \finaledit{I}ntelligence (AI) -- and young people are picking it up fast. Approximately 65\% of youth use generative AI tools \cite{salesforce2023new}, such as Large Language Models (LLMs), chatbots, and virtual assistants. Preliminary work suggests that AI-driven systems hold promise to build on the previously described benefits of online interactions by providing accessible, convenient, and personalized support \cite{haque2023overview}. Young people recognize this promise, as more than two-thirds of them have experience with AI-informed mental health apps \cite{gotzl2022artificial}, and many view AI as an acceptable solution to offer mental health support \cite{koulouri2022chatbots}.} \finaledit{Furthermore, research suggests that users not only trust AI systems for mental health support but also have a positive experience using them \cite{shan2022public}. For example, interventions using AI to deliver psychological support specifically for young people are suggesting that AI could be a feasible, engaging, and effective solution \cite{fitzpatrick2017delivering, frazier2016internet,fulmer2018using}.}

SIGCHI researchers~\cite{razi2023aCSCW,razi_lets_2020,naslund2016future,armstrong2009patient}, have thoroughly investigated the significance of online peer support in the lives of youth, discovering that while peers would like to offer support, they sometimes do not know how to help or feel the disclosure was inauthentic~\cite{huh2023help}. Furthermore, peer support providers sometimes struggle to maintain boundaries to express their difficulties in writing supportive and empathic responses due to lack of role clarity~\cite{kemp2012}. 
The notion of education as a solution to the lack of expertise \edit{of support givers} was validated by a later study where researchers \cite{sharma2023human} developed an AI system called '\textit{Hailey'} which helped develop more empathetic responses to support seekers in a mental health peer support platform. AI does have many limitations, such as its inability to understand human nuances~\cite{bishop2021artificial}, ethical \edit{risks of misclassification for sensitive topics~\cite{razi2021human}, and potential algorithmic stigmatization and harms~\cite{Andalibi2023conceptualizing}}.
This has led to continued efforts to explore the potential of AI as a supplementary, rather than a primary, source of support for youth.
However, youth's norms around peer support and the extent to which AI can effectively address their unique support needs based on frequent support-seeking topics of youth 
remain areas of ongoing investigation \cite{fulmer2019artificial}.

\edit{Therefore, our goal is to understand how AI compares to the different types of support young people encounter online within the context of the topics that young people specifically seek support about.}
To address the aforementioned gaps in the literature, we ask the following research question\edit{s}: 

\edit{\textbf{(RQ1)} \textit{What are young people's preferences regarding help-giving responses created by humans with various help-giving skills compared to AI-generated responses?}} 

\edit{\textbf{(RQ2)} \textit{How do young people's preferences for help-giving responses vary depending on the topic (relationships, suicidal thoughts, self-expression, or physical health) of the initial help-seeking message?}}

To answer these question\edit{s}, we analyzed 622 survey responses from young people \edit{to understand} their preferences about pre-written responses from various sources (AI, therapist, adult mentor, and peer) to four help-seeking messages on topics \edit{that young people frequently seek support about (}relationships, suicidal thoughts, self-expression, and physical health\edit{) \cite{huh2023help}}. 
We found that young people preferred AI-generated responses about less sensitive topics (relationships, self-expression, and physical health). For more sensitive discussions (suicidal thoughts), participants preferred responses from the adult mentor.
Our study contributes to the Human-Computer Interaction (HCI) research community in several ways. This paper provides insights into how young people perceive AI-generated responses in peer support situations. Our results also identify the popular conversational topics among young people where AI-generated responses are, and are not, well received. The findings bring implications for designing AI-powered systems to safely and effectively help youth \edit{during online support interactions}.

\section{Related Work}


\subsection{Young People's \edit{Mental Health} and Online Social Support}
More than 450 million people around the world struggle with mental illness \cite{saraceno2002world}, and young people are disproportionately likely to be among those who are suffering \cite{jurewicz2015mental}. In a cross-country survey, 
the World Health Organization found that nearly one-third of college students tested positive for at least one mental health disorder \cite{auerbach2018world}. Furthermore, mental illnesses like depression, substance use disorders, and psychosis are some of the biggest risk factors for suicide \cite{braadvik2018suicide}, which is the second-leading cause of death for teens and young adults \cite{curtin2022provisional}. With professional treatment being less affordable and less accessible than ever before \cite{kanagaraj2020why}, internet interventions are emerging as a promising avenue for formal treatment \cite{cunningham2014internet} and informal peer support \cite{woodruff2007evaluation,Hartikainen2021safesexting}.

Social media platforms are some of the most-visited websites on the internet, with some sites getting an estimated 1.7 billion visits every 6 months \cite{statista2021}. 
Researchers \cite{ridout2018use} suggested that youth benefit from online support exchanges on social media platforms; for example, actively engaging with peers online about mental health concerns is associated with an increased likelihood of seeking formal mental health care \cite{ridout2018use}. Unique platform affordances, like anonymity, can help users feel more comfortable being honest about their experiences and feedback \cite{de2014mental}. Also, support exchange on social media can lead to increased quality of life, heightened perceived support, and increased self-efficacy \cite{uehara1990dual}.

\subsection{AI-Powered Mental Health Support}
AI systems are \finaledit{evolving quickly} and have the potential to transform many fields, including mental health \cite{folstad2017chatbots,brandtzaeg2018chatbots}. Generative AI systems, like ChatGPT \footnote{https://chat.openai.com/} and Google Bard \footnote{https://bard.google.com/}, are machine learning models that can create human-like responses for accessible, affordable, and individualized support \cite{singh2023artificial}. 
\edit{The time frame since these technologies became available is very short, but users have started to explore} various ways in which ChatGPT can support mental health and archived \edit{their findings} in non-peer-reviewed venues. 
For instance, researchers explored ChatGPT as a therapist assistant \cite{eshghie2023chatgpt}, where it can work as a companion for patients between therapy sessions, collecting information, listening to patients attentively, offering validation, and sharing potential coping strategies. Another study (again non-empirical, non-peer-reviewed) examined the possibilities of using ChatGPT for child mental health \cite{imran2023chat} \edit{with} medical education for children, translation and assisting in provider-child patient communication, and access to medical information. \cite{van2023artificial} and \cite{kizito2023chatgpt} \edit{discuss similar possibilities for} health condition\edit{s that} can benefit from anonymity in accessing information \cite{kizito2023chatgpt}.

Research about the outcomes of implementing AI systems for mental health is mixed. For example, an analysis of user review comments found that users had mostly positive experiences with AI chatbots for symptoms of anxiety and depression \cite{ahmed_thematic_2022}, and surveys of chatbot systems can suggest that they are useful reporting tools for in-person harms \cite{rita2021chatbot} and can even help to increase youth participation in societal activities \cite{vaananen2020civicbots}. 
However, Martinengo et al. examined dialogues with AI conversational agents (CAs) and found that CAs from app stores are \textit{not} fit for comprehensive suicide risk management. According to their findings, this is because AI systems follow a predefined protocol for suicidal thoughts that is only activated in response to specific phrases that suggest immediate danger, meaning they cannot identify when a conversation is moving towards a problematic topic. Another issue is that CAs cannot send an unsolicited follow-up message after the conversation concludes to ensure the user is safe \cite{martinengo2022evaluation}. 


\edit{Researchers highlighted the importance of better understanding} how AI can maximize its ability to support young people by assisting in conversations \cite{Shao2023empathetic}. For instance, Sharma et al. \cite{sharma2023human} experimented with an AI agent that provides just-in-time feedback to peer support providers in an online mental health support platform and found that AI can help peers increase empathy in their support-giving messages. \edit{These findings are promising for the role of AI in peer-to-peer interactions, but the study did not focus on young people and their specific needs.} 

\subsection{AI \edit{and Online Support for Young People}}
Researchers \cite{carlbring2023new} have investigated \edit{many aspects of AI-enabled} technologies, \edit{including} personalization, ethical considerations, recalling \edit{information from} previous sessions, and detecting body gestures and expressions \cite{carlbring2023new}. At the same time, researchers highlight the lack of accuracy, reliability, bias, and unpredictability of outputs of these generative AI systems \cite{mardikaningsih2023risk}, calling for appropriate mitigation strategies, awareness, and careful consideration of these challenges before jumping into experimenting the possibilities of generative AI on sensitive topics, such as high-risk mental health conversations (e.g., suicide ideation).


The common shortcomings of these existing works include empirical testing of how end users might perceive these generative AI-powered systems that support mental health. \edit{While previous work explored people's perceptions of AI-based agents for mental health \cite{haque2023overview, aggarwal2023artificial}, the verbal expressions and stylistic characteristics of more recent language-based AI technologies (e.g., LLMs) have more human-like characteristics than other AI agents to date \cite{pan2023llms}, which requires furture investigation. Given this technology is new and recent, researchers have only begun to understand the nuanced perceptions of how people over-trust, under-trust, or assess the evolving nature of AI agents' abilities. Additionally, there is very little research on how specific populations, such as youth, might react to these technologies.}

\edit{The critical role that online social support plays for youth, the advent of new AI technologies that have great potential to support mental health, and the lack of understanding around young people's perceptions of more advanced AI agents place our study in a unique position to contribute to the gaps noted in the literature. Our study is the first to investigate} young people's perceptions of how helpful, inappropriate, and problematic AI-generated responses are compared to multiple human-generated responses. These tests were conducted blinded, thus allowing us to objectively evaluate initial reactions to the style of the responses themselves rather than young people's preconceptions around AI. 

\edit{In summary, young people often seek support online, especially for mental health concerns \cite{bae2021social, pretorius2019young}. That support can come from individuals with varying degrees of expertise and experience, and research has shown that these interactions are beneficial \cite{pretorius2019young}. Now, AI systems are permeating digital conversations, but we do not know how young people feel about the content these systems produce. To address this gap, we studied how young people respond to AI-generated messages compared to the different types of support young people encounter online within the context of the topics that young people specifically seek support about.}

\section{Methods}
\label{sec:methods}
Using an online survey, we examined how young people react to different human- and AI-generated responses to help-seeking messages. In this section, we describe our \edit{participant criteria and recruitment process}, explain our survey design, \edit{review ethical considerations}, and describe our approach to data analysis. 


\subsection{Participant Recruitment \edit{Criteria and} Process}

\edit{Our target population was youth aged 18-24. While youth includes individuals aged 10-24 \cite{orenstein_eriksons_2022}, as it is a time in which physiological and psychosocial development impact well-being in ways that drastically extend the transition period to adulthood \cite{sawyer_age_2018}, we intentionally did not recruit minors. This is because of the sensitive nature of some topics in this study, and we were cautious to avoid unnecessarily exposing minors to topics like suicidal thoughts.}

\edit{Based on a population of youth in the U.S. of 31 million \cite{annie2023young}, we determined we needed a sample of at least 385 participants for a 95\% confidence level with a 5\% margin of error and 664 participants for a 99\% confidence level with a 5\% margin of error. Our target was a minimum of 385, but we wanted to get as close to 664 as possible.}

We recruited participants by sharing a digital flyer \edit{on Instagram, Facebook, X (formerly known as Twitter), Discord, and Slack. The flyer included a QR code linked to} an online screening survey to determine prospective participants' eligibility, which required that they were 18 to 24 years old, currently living in the U.S., fluent in English, and have stable access to the Internet. Eligible respondents (N=1,096) continued to the IRB-approved study information sheet, which they were asked to review before continuing to the main survey. To ensure the quality of the responses, there was a built-in attention check question \edit{halfway through the main survey} that asked participants to accurately indicate which help-seeking message topic we had previously asked about.
After finishing the survey, participants could provide their contact information to receive \$5 as payment for their time. We received 866 survey responses from eligible participants, of which 622 were deemed fit for analysis \edit{(see table \ref{tab:participant-demographics} for demographic information)} because they were complete and passed the built-in attention check question.

\subsection{Survey Design}
\label{survey-design}
\edit{We used a 4 x 4 study design with four help-seeking messages and four pre-written responses (henceforth referred to as "responses") to each help-seeking message (4 help-seeking messages x 4 responses = 16 total combinations). For each combination, we asked multiple questions. This section provides more details about the help-seeking messages, responses, and survey structure (see Figure \ref{fig:methods-horizontal}).}

\subsubsection{\edit{Help-Seeking Messages and Pre-Written Responses}}
\edit{The help-seeking messages came from \cite{huh2023help}, in which the authors analyzed direct message conversations between youth and categorized youth private messages for support into four main themes: (1) relationships, (2) suicidal thoughts, (3) self-expression, and (4) physical health. To capture these themes in our survey, we used the full help-seeking messages that the excerpts in \cite{huh2023help} came from.}


The responses were written by four different sources: a (1) peer, (2) adult mentor, (3) therapist, and (4) \finaledit{A}rtificial \finaledit{I}ntelligence (AI).
\edit{The} \textbf{peer} responses \edit{in our survey, like the help-seeking messages, came directly from the dataset used in \cite{huh2023help} and were the real responses youth sent after receiving the corresponding help-seeking messages from \cite{huh2023help}. For the \textbf{adult mentor} responses, we wanted the message to reflect best practices for non-clinical crisis interventions. To do this, we identified an intensive, evidence-based training program through Crisis Text Line\footnote{\url{https://www.crisistextline.org}}, a global not-for-profit organization that provides mental health support via text. Given the existing research to support that this organization’s practices are effective \cite{crisis2023decade,gould2022crisis}, the second author used the training materials to generate the adult mentor’s responses to capture the appropriate tone through deliberate phrasing.}
To capture evidence-based professional training, we had a licensed clinical psychologist who also runs an APA-accredited clinical training program write the \textbf{therapist} responses. 
We used responses generated by ChatGPT for the \textbf{AI} responses. To understand how AI and LLMs could advise and support young people, we prompted ChatGPT, a popular AI-powered chatbot used by many young adults \cite{vogels2023majority}, in May 2023 (GPT-3.5, free version) by asking it to write a response to each help-seeking message ``in the appropriate style and of the appropriate length for a mobile text message" \edit{(see Appendix \ref{appendix-context} for the full context of the prompts we used)}. We chose this prompt after exploration with ChatGPT and some prompt engineering. Initial attempts to generate a response by providing only the help-seeking message repeatedly generated responses in the form of a long bulleted list of advice. Compared to our earlier exploration, ChatGPT's response to our most current prompt was considerably shorter and involved more interaction with the help seeker, which was suitable for a text message. \edit{The full text of the help-seeking messages and responses are in appendices \ref{appendix-messages} and \ref{appendix-responses}, respectively.}




Participants were blinded about where the responses came from because we wanted to understand perceptions of the content as part of peer support exchange, not feelings about different sources of support engaging in the exchange. We also randomized both the order in which the scenarios were presented and the individual questions about each response to prevent question order bias.  

\subsubsection{\edit{Survey Structure}}
\edit{At the beginning of the survey, we asked participants how often they seek and give support online. We used two multiple-choice questions, one for seeking support and one for giving support, to which the answer choices were never, rarely, sometimes, often, or always.}

\edit{Next, we presented the sixteen help-seeking message x response combinations in four sections; each section contained one help-seeking message and its four corresponding responses. Each section had the same structure. First, we instructed participants to imagine they received the help-seeking message from a friend as a text or direct message on social media. Then, we} asked participants to rate the extent to which they agreed with the following statements on a 5-point Likert scale (1 = not at all, 5 = extremely): "I like this response," "I would send this response," "This response contains helpful advice," "This response is problematic," and  "This response is inappropriate." \edit{At the end of each section,} we asked participants to type a brief text response about \edit{why they did and did not like certain combinations.}

\edit{The rationale in using the Likert scale response options was to measure the direction and intensity of the respondent’s attitude and preference toward the specific message. If attitude and preference are measured in a binary fashion (i.e., yes or no), then the data is not robust enough to capture the intensity of the attitude and preference. Because we asked respondents about their attitude and preference for each help-seeking message x response combination, it was important to capture the relative intensity of their attitude and preference for each message to allow for nuance within the data, resulting in a more accurate and representative understanding of feedback. Seeing as a limitation of Likert scale responses is not understanding the "why'' behind an attitude or preference, we then followed up with the qualitative question.} 

\edit{An attitude or a preference is a psychological construct. In our study, the attitude or preference we measured represent the survey respondent’s predisposition to respond favorably or unfavorably to the scenario (i.e., “I like this response”). Attitudes and preferences are often - although not always - considered precursors to behavior. Numerous studies have demonstrated that individuals often express their attitudes but do not always act in accordance with those attitudes (\cite{bechler2021attitude}). In our study, we used several additional survey questions to examine whether there was consistency with what the respondents liked and what they found helpful (i.e., “I would send this response” and “This response contains helpful advice”). There is a compelling body of research in social psychology that suggests there are times of conflict between what people want (immediate gratification, for example) and what they believe they should do or what they find most helpful (longer-term benefits; \cite{bitterly2017risky}).}

\edit{We included opposing statements for the Likert scale questions for multiple reasons. First, } we know from psychology literature \cite{dakof1990victims} that people may not always seek help or solutions when they discuss problems or concerns; instead, they may primarily seek emotional support, active listening, or validation. 
Therefore, these statements were designed to capture those preferences and validate the cohesiveness in survey responses (i.e., it is unlikely that a participant finds a response both helpful and inappropriate). \edit{We also measured opposing concepts because literature in survey design suggests that participants can present acquiescence bias, which is the tendency to agree with statements regardless of the content. Also, social desirability bias would make participants inclined to answer in ways they believe are more socially acceptable or favorable. Accordingly, asking about opposing concepts to force participants to think more critically about their answers would increase the quality of survey responses \cite{kuru2016improving}.}

\begin{figure*}[h!]
    \centering
    \includegraphics[scale = 0.7]{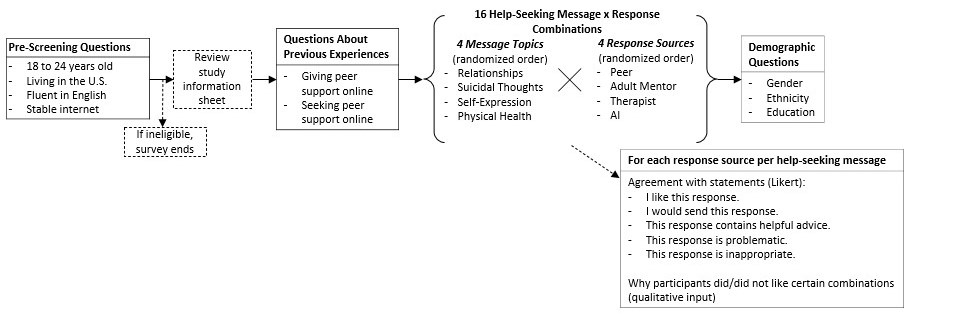}
    \caption{\edit{Visualization of the survey structure described in section 3.2.2}}
    \Description{The figure shows the survey structure, which included pre-screening questions, a study information sheet, questions about previous experiences, sixteen help-seeking message and response combinations with questions about each one, and demographic questions.}
    \label{fig:methods-horizontal}
\end{figure*}

We also asked for our participants’ demographic information (Table\ref{tab:participant-demographics}). \edit{See Figure \ref{fig:methods-horizontal} for a visualization of the survey structure.}


\begin{table}[h!]
\centering
\small
\begin{tabular}{|ccc|}
\hline
\multicolumn{1}{|c|}{\textit{\textbf{Variable}}} & \multicolumn{1}{c|}{\textit{\textbf{Count (N)}}} & \textit{\textbf{Percent of total (\%)}} \\ \hline
\multicolumn{1}{|c|}{All Responses} & \multicolumn{1}{c|}{622} & 100.00\% \\ \hline
\multicolumn{3}{|c|}{\textbf{Gender}} \\ \hline
\multicolumn{1}{|c|}{Male} & \multicolumn{1}{c|}{403} & 64.79\% \\ \hline
\multicolumn{1}{|c|}{Female} & \multicolumn{1}{c|}{216} & 34.73\% \\ \hline
\multicolumn{1}{|c|}{Non-binary} & \multicolumn{1}{c|}{2} & 0.32\% \\ \hline
\multicolumn{1}{|c|}{Prefer to self-identify} & \multicolumn{1}{c|}{1} & 0.16\% \\ \hline
\multicolumn{1}{|c|}{Prefer not to say} & \multicolumn{1}{c|}{0} & 0.00\% \\ \hline
\multicolumn{1}{|c|}{Total} & \multicolumn{1}{c|}{622} & 100.00\% \\ \hline
\multicolumn{3}{|c|}{\textbf{Ethnicity}} \\ \hline
\multicolumn{1}{|c|}{Caucasian} & \multicolumn{1}{c|}{295} & 47.43\% \\ \hline
\multicolumn{1}{|c|}{African American} & \multicolumn{1}{c|}{193} & 31.03\% \\ \hline
\multicolumn{1}{|c|}{Native American} & \multicolumn{1}{c|}{34} & 5.47\% \\ \hline
\multicolumn{1}{|c|}{Latino or Hispanic} & \multicolumn{1}{c|}{26} & 4.18\% \\ \hline
\multicolumn{1}{|c|}{Mixed Race} & \multicolumn{1}{c|}{24} & 3.86\% \\ \hline
\multicolumn{1}{|c|}{Asian} & \multicolumn{1}{c|}{17} & 2.73\% \\ \hline
\multicolumn{1}{|c|}{Multiple} & \multicolumn{1}{c|}{16} & 2.57\% \\ \hline
\multicolumn{1}{|c|}{Prefer not to say} & \multicolumn{1}{c|}{9} & 1.45\% \\ \hline
\multicolumn{1}{|c|}{Native Hawaiian or Pacific Islander} & \multicolumn{1}{c|}{6} & 0.96\% \\ \hline
\multicolumn{1}{|c|}{Other} & \multicolumn{1}{c|}{2} & 0.32\% \\ \hline
\multicolumn{1}{|c|}{Total} & \multicolumn{1}{c|}{622} & 100.00\% \\ \hline
\multicolumn{3}{|c|}{\textbf{Education}} \\ \hline
\multicolumn{1}{|c|}{Some high school} & \multicolumn{1}{c|}{17} & 2.73\% \\ \hline
\multicolumn{1}{|c|}{High school or equivalent} & \multicolumn{1}{c|}{147} & 23.63\% \\ \hline
\multicolumn{1}{|c|}{Trade school} & \multicolumn{1}{c|}{100} & 16.08\% \\ \hline
\multicolumn{1}{|c|}{Bachelor's degree} & \multicolumn{1}{c|}{297} & 47.75\% \\ \hline
\multicolumn{1}{|c|}{Master's degree} & \multicolumn{1}{c|}{39} & 6.27\% \\ \hline
\multicolumn{1}{|c|}{Doctoral degree} & \multicolumn{1}{c|}{12} & 1.93\% \\ \hline
\multicolumn{1}{|c|}{Other} & \multicolumn{1}{c|}{3} & 0.48\% \\ \hline
\multicolumn{1}{|c|}{Prefer not to say} & \multicolumn{1}{c|}{7} & 1.13\% \\ \hline
\multicolumn{1}{|c|}{Total} & \multicolumn{1}{c|}{622} & 100.00\% \\ \hline
\end{tabular}
\caption{Participant demographics}
\label{tab:participant-demographics}
\end{table}

\subsection{Ethical Considerations}
\finaledit{Ethical research conduct is of paramount importance, particularly with regard to sensitive topics.} This study \edit{was submitted for full-review} to Drexel University's IRB \edit{and} was approved as an exempt study\edit{. This} meant participants were presented with the study information sheet but not required to sign a consent document. \edit{Despite our study's exempt status, we took multiple precautions to address potential risks} \finaledit{and maintain emotional safety for participants, especially} \edit{due to the sensitive nature of our survey materials.} \finaledit{We specifically and intentionally included a board-certified clinical psychologist on our research team to carefully review all language in the survey questions and AI-generated responses to minimize potential risks and monitor incoming data for concerning submissions.} \edit{We also presented all participants with a study information sheet that contained the same information as a full consent document, including detailed descriptions of} the research objective, the survey procedures involved, the voluntary nature of the study, the potential risks involved, the anonymization process, and incentives. \edit{In lieu of a consent signature, all participants had to acknowledge they read this document before beginning any study procedures.}

\edit{To protect participants' anonymity and confidentiality, o}nly the research team members specified in the IRB protocol were granted access to the data. \finaledit{The researchers stored and accessed all data on a university-approved server. This study took place entirely online, meaning participants could complete the survey in a private location of their choice.} \edit{When participants finished their participation, they were each assigned a unique identifier, known only to the researchers on the IRB protocol, that we used to identify the participants instead of their names. The PIs maintained an Excel spreadsheet containing participants’ contact information, which we did not use in our data analysis or any publication resulting from this research.}

\subsection{Data Analysis}
To ensure we collected high-quality data, we first looked at the correlations between how much participants 'liked' the responses, felt they `would send' the message to the help-seeker, and how `helpful' they thought it was compared to how `problematic' and `inappropriate.' We did this because it is possible for participants to prefer a response for paradoxical reasons, such as finding the content comically inappropriate - fortunately, this was not the case in our data. Using means across the opposing attributes, we confirmed that the responses from sources that participants preferred more, said they were more likely to send, and felt were more helpful were also perceived as less inappropriate and less problematic. 
Participants also tended to feel responses were more helpful than inappropriate or problematic, which means we avoided doing more harm than good, as is essential for ethical research.

We also wanted to understand how participants reacted to the different responses beyond which one they `liked' the most (i.e., preferred). To do this, we looked at aggregate responses for all help-seeking messages by the source to understand how helpful the responses were (`helpful'), how likely they were to send them (`would send'), and how inappropriate (`inappropriate') and problematic (`problematic') they felt the responses were. In subsequent sections and figures, we refer to these four measures as `attributes' of participants' responses. \edit{Furthermore, the language we use to describe the findings for the 'would send' attribute includes the phrase “more likely.'' This refers to the extent to which participants agreed they would send the response in question, not that we performed probability or likelihood analysis as that was not part of our study methodology.}

\edit{We analyzed the quantitative data using R software.} \finaledit{Our group size (N) for all statistical tests was equal to our total number of participants, which was 622.} \edit{Before comparing means across combinations, we used Bartlett's test of homogeneity to determine whether variance was equal} \finaledit{and normality of the data distribution.} \edit{When Bartlett's test was significant, indicating that variance is not equal across variables, we used Welch one-way tests for nonparametric data to compare means. When Barlett's test was not significant, meaning variance is equal across variables, we used a one-way ANOVA. We report the p-value for Bartlett's test before all F-statistics to signal which test we used.}

We analyzed the qualitative responses using reflexive thematic analysis \cite{braun2019reflecting} \edit{because it allows appropriate flexibility for inductive, iterative analysis to draw rich conclusions. We took an inductive approach to ensure conceptual coherence because our goal was to identify and develop themes directed by the content of our data, not by existing concepts or ideas. Rather than focusing on specific questions, our objective was to complement our quantitative findings and further explore reasons why participants did, or did not, like specific responses to help-seeking messages about relationships, suicidal thoughts, self-expression, and physical health. Since we conducted this qualitative analysis after analyzing participants' quantitative data, our analytical framework was situated within an essentialist perspective; in other words, we presupposed that participants' open-ended responses reflected their opinions expressed in their previous quantitative survey responses.}

\edit{To start, the second and third authors familiarized themselves with the data by reviewing all qualitative responses and taking notes along the way. They used those notes to identify initial themes for discussion and to share with the rest of the research team. The research team met multiple times to discuss and iterate over the themes and relevant examples to produce the findings in the results section of this paper. We did not use a traditional codebook because the qualitative analyses complement our quantitative findings rather than act as a stand-alone analysis. The first, second, and third authors identified the examples used in our findings section to illustrate our main points.}

\section{Results}

Next, we describe participants' preferences for and reactions to different support-giving responses (peer, adult mentor, therapist, and AI) across all help-seeking topics and how these preferences and reactions vary depending on the topic of the help-seeking message they are responding to (relationships, suicidal thoughts, self-expression, or physical health).

\edit{Almost all of our participants (97\%) reported at least some previous experience seeking and giving support online (i.e., they did not select "never"). The relative frequency of responses was the same for both seeking and giving support. The most frequent response was "sometimes" (seeking: n = 277, 45\%; giving: n = 238, 38\%), second was "often" (seeking: n = 148, 24\%; giving: n = 174, 28\%), third was "rarely" (seeking: n = 113, 18\%; giving: n = 137, 22\%), fourth was "always" (seeking: n = 65, 10\%; giving: n = 57, 9\%), and least frequent was "never" (seeking: n = 19, 3\%; giving: n = 16, 3\%).}

\subsection{Young People's Preferences Regarding Human and AI-generated Help-Giving Responses (RQ\edit{1})}
We found that the human-generated adult-mentor response performed best when we combined responses across all help-seeking topics and compared the average responses from the four sources (peer, adult mentor, therapist, and AI).
How much participants liked each source's responses were significantly different (\edit{Bartlett's test: p < 0.05;} \textit{F}(3.0,5517.7) = 99.20, p < 0.001) across all help-seeking messages. \edit{There were also significant differences in the extent to which participants found the responses helpful (Bartlett's test: p < 0.05; \textit{F}(3.0,5519.6) = 104.37, p < 0.001), problematic (Bartlett's test: p < 0.05; \textit{F}(3.0,5525.6) = 20.159, p < 0.001), inappropriate (Bartlett's test: p > 0.05; \textit{F}(3.0,9948) = 19.452, p < 0.001), and how likely they were to send the response (Bartlett's test: p < 0.05; \textit{F}(3.0,5513.9) = 91.09, p < 0.001).} 

The post-hoc analysis showed that participants liked the adult mentor's responses the most (M=3.51, SD=0.02), followed by AI (M=3.46, SD=0.02), then the therapist's (M=3.29, SD=0.02), and the peer responses were their least favorite (M=3.03, SD=0.02). This means that before considering the topic of the help-seeking message, it seemed like humans, specifically adult mentors, outperformed AI in generating help-giving responses, although peer responses were not as liked by the participants. However, such responses varied after considering help-seeking topics, which we will discuss. 
Participants felt the adult mentor's responses were the most helpful (M=3.50, SD=0.99) of the responses presented and also indicated that they were most likely to send the adult mentor's responses to the help-seeker (M=3.48, SD=0.98), followed by AI's responses (how helpful: M=3.49, SD=1.05; \edit{would send}: M=3.42, SD=1.12), then the therapist's responses (how helpful: M=3.29, SD=1.07; \edit{would send}: M=3.25, SD=1.12), and the peer's responses receiving the lowest score for those preferences (how helpful: M=3.02, SD=1.15; \edit{would send}: M=3.00, SD=1.18).

\begin{figure*}[!t]
    \centering
    \includegraphics[scale = 0.65]{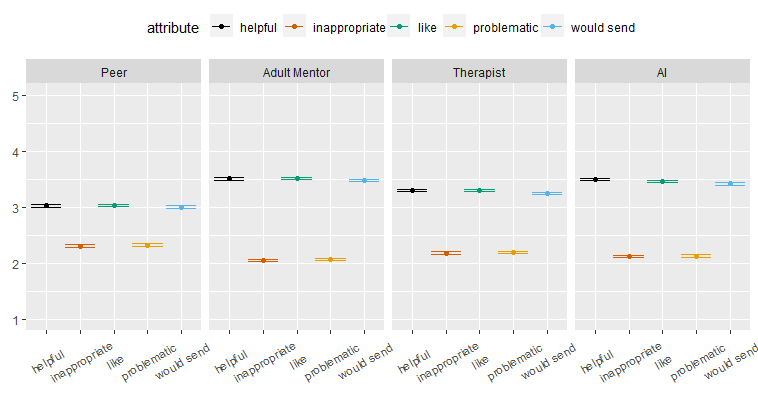}
    \caption{Participant preferences \edit{for human and AI-generated responses} by source \edit{(RQ1)}. The y-axis is how much participants agreed with the different attributes of their reaction \edit{(i.e., how strongly they agreed with the statements described in section 3.2.2). Points represent means, and the bars on either end are standard deviation.}}
    \Description{The figure includes four dot plots with error bars for data from all four scenarios combined. Each plot corresponds to a different response source (peer, adult mentor, therapist, and AI). The x-axis lists different attributes of participants' reactions to help-giving responses (helpful, inappropriate, like, problematic, and would send) and the y-axis represents how much participants agreed with each attribute of their reaction (i.e., how strongly they agreed with the statements described in section 3.2.2).}
    \label{fig:source-reactions}
\end{figure*}

How problematic and inappropriate participants felt the responses were, depending on the source of the response, followed the opposite pattern. The peer responses were perceived to be the most problematic and inappropriate (how problematic: M=2.32, SD=1.24; how inappropriate: M=2.30, SD = 1.24), followed by the therapist's responses (how problematic: M=2.19, SD=1.22; how inappropriate: M=2.18, SD=1.23), AI's responses (how problematic: M=2.13, SD=0.02; how inappropriate: M=2.12, SD=0.02), and the adult mentor's responses as the least problematic (M=2.07, SD = 0.02) and inappropriate (M=2.05, SD = 0.02). See Figure \ref{fig:source-reactions} for a visualization of means and standard deviations for participants' reactions to the responses by source.

The findings so far suggest that participants regarded AI-generated responses highly, above even therapist and peer responses, although adult mentors' responses were considered most helpful and most likely to be sent in response to the help-seeking messages. 

\subsection{Young People's Preferences for Help-Giving Responses Depending on the Topic of the Initial Help-Seeking Message and Rationale Behind Them (RQ\edit{2})}

\subsubsection{Relationships}
\label{sec:relationships}
\begin{figure*}[hbt!]
    \centering
    \includegraphics[scale=0.65]{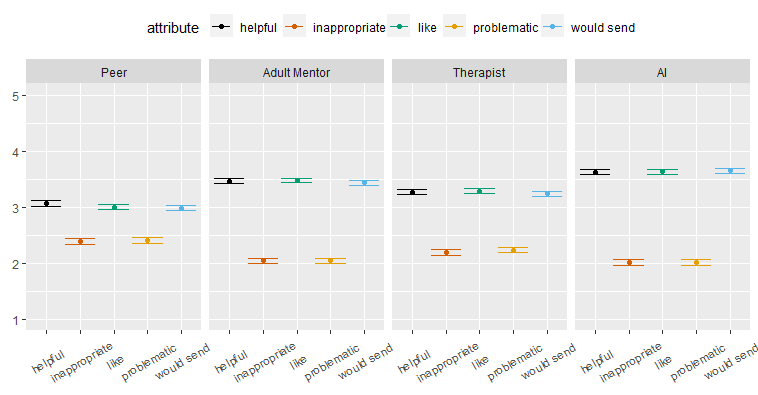}
    \caption{\edit{Participant preferences for human and AI-generated responses} by source for help-seeking message about relationships \edit{(RQ2, relationships)}.
    \edit{The y-axis is how much participants agreed with the different attributes of their reaction (i.e., how strongly they agreed with the statements described in section 3.2.2). Points represent means, and the bars on either end are standard deviation.}}
    \Description{The figure includes four dot plots with error bars for data from the scenario about relationships. Each plot corresponds to a different response source (peer, adult mentor, therapist, and AI). The x-axis lists different attributes of participants' reactions to help-giving responses (helpful, inappropriate, like, problematic, and would send) and the y-axis represents how much participants agreed with each attribute of their reaction (i.e., how strongly they agreed with the statements described in section 3.2.2).}
    \label{fig:plot-relationships}
\end{figure*}

\edit{How much participants liked each source's responses to the help-seeking message about relationships were significantly different (Bartlett's test: p < 0.05; \textit{F}(3.0,1378) = 39.145, p < 0.001) as were the extent to which participants found the responses helpful (Bartlett's test: p < 0.05; \textit{F}(3.0,1377.4) = 33.72, p < 0.001), problematic (Bartlett's test: p < 0.05; \textit{F}(3.0,1379.0) = 14.47, p < 0.001), inappropriate (Bartlett's test: p > 0.05; \textit{F}(3.0,2484) = 12.113, p < 0.001), and how likely they were to send the response (Bartlett's test: p < 0.05; \textit{F}(3.0,1375.6) = 43.24, p < 0.001).}
Participants preferred the response about relationships that AI-generated more than any of the human-generated options.
They liked AI's response the most (M=3.63, SD=0.04) and thought it was the most helpful (M=3.63, SD=1.00), the least problematic (M=2.01, SD=1.16) or inappropriate (M=2.01, SD=1.20), and said they were most likely to send the AI's response to the help-seeker (M=3.65, SD=0.96). On the other hand, participants did not like the peer's response about relationships (M=3.00, SD=1.15) and felt it was the most problematic (M=2.41, SD=1.24) and inappropriate (M=2.38, SD=1.24), and least helpful (M=3.07, SD=1.13); consequently, they were also least likely to send the peer's response (M=2.98, SD=1.17). See Figure \ref{fig:plot-relationships} for a visual comparison of the means and standard deviations for reactions to all responses about relationships. 

This sentiment was reflected in open-ended questions, where participants shared they liked the balance and empathy in the \textbf{AI-generated} responses. They felt the AI-generated response invited the help-seeker to share additional information without being overly intrusive or demanding. As one participant put it, the response was \textit{``helpful...
[and] also supportive and feels like a safe space.''} (P453, Female).
Some participants described anthropomorphic reasons, such as the AI-generated response seemed like it came from someone who is a good friend: \textit{``They are really trying to help the friend so much and we all need such friends.''} (P434, Female). There were also mentions of qualities participants associated with the AI response, including that they thought it was \textit{``[empathetic]...supportive and caring.''} (P46, Male). Even though participants did favor the AI-generated response to the help-seeking message about relationships, they did note some room for improvement. Namely, the response doesn't consider the entire context of the situation and \textit{``accuses the ex without listening to his side of the story.''} (P118, Male).

Participants noted that the
\textbf{adult mentor's responses} also provided acknowledgment and validation, but it was potentially less useful and lacked clear direction. Those who liked this response noted that \textit{``it is respectful and...made [them] feel like the person understood [the help-seeker's] feelings.''} (P50, Male). Participants also noted a sense of maturity in how the response did not judge the help-seeker: \textit{``...the [response] answered the question in a mature way by making the victim not feel offended by going back to the ex. The response is just showing him or her that is what we [humans] go through in terms of love life.''} (P403, Male). Despite these positive attributes of the adult mentor's response, some participants felt it lacked concrete advice and did not give specific suggestions for what to do next. As mentioned in the previous paragraph, participants liked that the AI-generated response guided the help-seeker to provide additional detail in a gentle way - they did not feel the adult mentor's response had this same quality. Instead, participants felt the adult mentor's response was, as one participant said, \textit{``
beating around the bush''} (P341, Male) and \textit{``
didn't provide a way of how to solve the situation clearly.''}

Participants liked the \textbf{therapist's response} because they felt it was extremely understanding and non-judgmental but noted that it seemed to delay giving actionable advice. Even though they did not know a therapist wrote one of the responses, they picked up on the professional tone: \textit{``This person provides professional remedy and ... [asks] the victim to share their mixed emotions and from there she/he will be in a position to offer strategies.''} (P481, Female). They saw the response as \textit{``very empathetic without shaming the friend''} (P587, Female), but noted that it felt like \textit{``scheduling [giving] help to next time when the [help-seeker] is in need of it now.''} (P503, Female). They appreciated the supportive tone but would have liked more concrete and immediate advice rather than beating around the bush and delaying support.

We received the most negative comments about the \textbf{peer's response} because participants felt it was terse, suggested hasty actions, and was not encouraging. While some appreciated the response's brevity, the way it was written came across as \textit{``too direct with no empathy.''} (P163, Male). This lack of empathy made participants feel like the sender had a bad attitude and was being too tough on the help-seeker: \textit{``It sounds harsh and rude, and also it doesn't consider the feelings of the victim.''} (P452, Female).

\subsubsection{Suicidal Thoughts}
\label{sec:suicidal-thoughts}
\edit{How much participants liked each source's responses to the help-seeking message about suicidal thoughts were significantly different (Bartlett's test: p < 0.05; \textit{F}(3.0,1375.3) = 64.08, p < 0.001) as were the extent to which participants found the responses helpful (Bartlett's test: p < 0.05; \textit{F}(3.0,1375.6) = 59.14, p < 0.001), problematic (Bartlett's test: p > 0.05; \textit{F}(3.0,2484.0) = 20.40, p < 0.001), inappropriate (Bartlett's test: p > 0.05; \textit{F}(3.0,2484) = 23.46, p < 0.001), and how likely they were to send the response (Bartlett's test: p < 0.05; \textit{F}(3.0,1372.9) = 73.75, p < 0.001).}

\begin{figure*}[hbt!]
    \centering
    \includegraphics[scale=0.65]{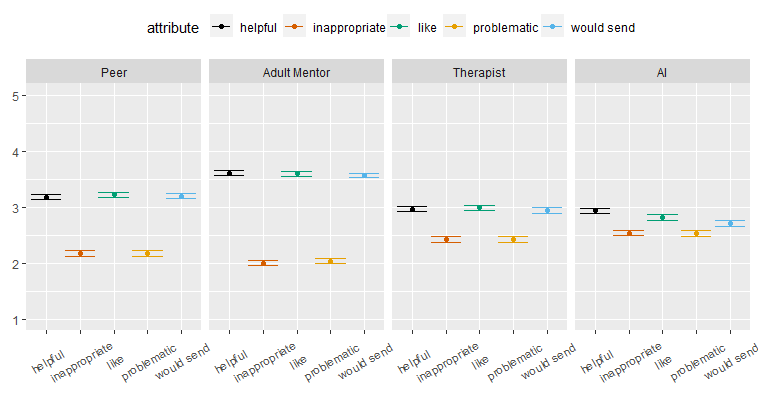}
    \caption{\edit{Participant preferences for human and AI-generated responses} by source for help-seeking message about suicidal thoughts \edit{(RQ2, suicidal thoughts)}. \edit{The y-axis is how much participants agreed with the different attributes of their reaction (i.e., how strongly they agreed with the statements described in section 3.2.2). Points represent means, and the bars on either end are standard deviation.}}
    \Description{The figure includes four dot plots with error bars for data from the scenario about suicidal thoughts. Each plot corresponds to a different response source (peer, adult mentor, therapist, and AI). The x-axis lists different attributes of participants' reactions to help-giving responses (helpful, inappropriate, like, problematic, and would send) and the y-axis represents how much participants agreed with each attribute of their reaction (i.e., how strongly they agreed with the statements described in section 3.2.2).}
    \label{fig:plot-suicidal-thoughts}
\end{figure*}

The help-seeking topic of suicidal thoughts was the only one for which all the human-generated responses (especially the one from the adult mentor) outperformed  AI's response. Participants liked the adult mentor's response the most (M=3.59, SD=0.95) and felt it was the most helpful (M=3.61, SD=0.94), least problematic (M=2.04, SD=1.18), least inappropriate (M=2.00, SD=1.20), and that they were most likely to send it to the help-seeker (M=3.56, SD=0.94). They also least liked AI's response (M=2.82, SD=1.19) and saw it as the most problematic (M=2.53, SD=1.26), most inappropriate (M=2.54, SD=1.27), least helpful (M=2.93, SD=1.12) and identified it as the one they were least likely to send to the help-seeker (M=2.71, SD=1.21).
See Figure \ref{fig:plot-suicidal-thoughts}. 

When describing why they preferred the \textbf{adult mentor's response} to this sensitive topic, participants noted that they liked how it conveyed genuine concern by inviting the help-seeker to share more details and put effort into providing more support. \textit{``There’s a feeling of relating with the individual here. There’s a feeling of care and [that they're] really interested and [want] to understand the individual}
\textit{[and their] situation.''} (P124, Male).
The only area of improvement that participants mentioned for the adult mentor response is that the message could have been a little shorter, but did not note what content should be cut. 

The \textbf{peer's} response was the second most preferred, and participants felt it was superficial, inconsiderate, and did not sufficiently acknowledge the help-seeker's struggles. Interestingly, participants likened the peer response to an AI-generated response, saying it sounded \textit{``...typical of a bot''} (P478, Female). Unlike the adult mentor's response, participants did not like how the peer's response was \textit{``so shallow [and] does not give the victim...much [detail] on how to tackle the situation and [get] helped.''} (P403, Male). 

Participants noted that the \textbf{therapist's response} immediately asked for more information in an attempt to ensure the help-seeker's safety - not knowing that this message was written by a clinician, some thought this sounded \textit{``professional''} while many others took it as intrusively asking for more information right away instead of offering support. \textit{``I think this response makes the person in need sink deeper into her trouble rather than pull her out. It's a bit inappropriate.''} (P451, Female). Since the initial help-seeking message did not use the word "suicide," many participants also did not like that the therapist's response immediately drew that conclusion and felt that calling it out could come across as judgemental: \textit{``I hate the assumption about self harm and suicide. It feels really I'm going to insert myself into your life to fix it and doesn't feel friendly at all.''} (P621, Non-binary).

The \textbf{AI-generated} response overall performed the worst for the help-seeking topic about suicidal thoughts. While some appreciated that the response was \textit{``actually truthful about the person's ability''} (P52, Male), most thought the referral to a healthcare provider conveyed less desire to help in that moment. \textit{``Fine, you can’t provide help but that’s not the best way to put it. You could always have the right choice of words and making the right sentence in this situation before suggesting a mental health professional and not everybody will like the fact that they are referring to them as going crazy. Not really cool though''} (P124, Male).

\subsubsection{Self-Expression}
\label{sec:self-expression}
\edit{How much participants liked each source's responses to the help-seeking message about self-expression were significantly different (Bartlett's test: p < 0.05; \textit{F}(3.0,1378.4) = 37.62, p < 0.001) as were the extent to which participants found the responses helpful (Bartlett's test: p < 0.05; \textit{F}(3.0,1378.4) = 39.53, p < 0.001), problematic (Bartlett's test: p > 0.05; \textit{F}(3.0,2484.0) = 4.88, p < 0.001), inappropriate (Bartlett's test: p > 0.05; \textit{F}(3.0,2484) = 5.56, p < 0.001), and how likely they were to send the response (Bartlett's test: p < 0.05; \textit{F}(3.0,1377.4) = 34.60, p < 0.001).}

\begin{figure*}[hbt!]
    \centering
    \includegraphics[scale=0.65]{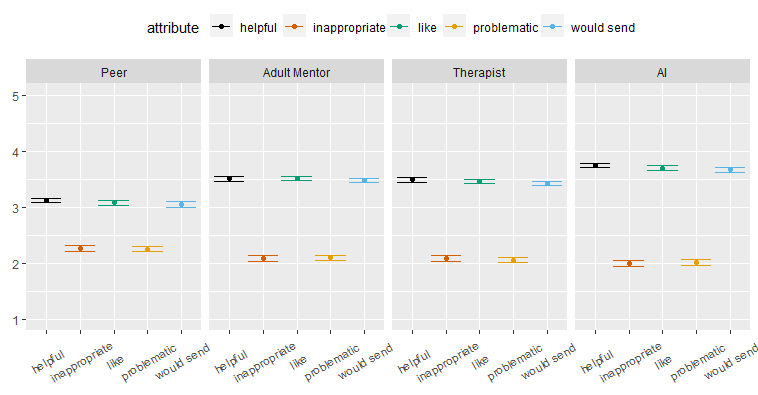}
    \caption{\edit{Participant preferences for human and AI-generated responses} by source for help-seeking message about self-expression \edit{(RQ2, self-expression)}. \edit{The y-axis is how much participants agreed with the different attributes of their reaction (i.e., how strongly they agreed with the statements described in section 3.2.2). Points represent means, and the bars on either end are standard deviation.}}
    \Description{The figure includes four dot plots with error bars for data from the scenario about self-expression. Each plot corresponds to a different response source (peer, adult mentor, therapist, and AI). The x-axis lists different attributes of participants' reactions to help-giving responses (helpful, inappropriate, like, problematic, and would send) and the y-axis represents how much participants agreed with each attribute of their reaction (i.e., how strongly they agreed with the statements described in section 3.2.2).}
    \label{fig:plot-self-expression}
\end{figure*}

Participants' reactions to the responses about self-expression followed a similar pattern to the responses about relationships, meaning AI's response outperformed all the human-generated responses, especially the peer's response. We found that participants liked AI's response the most (M=3.70, SD=0.98), felt it was the most helpful (M=3.75, SD=0.94), said they were most likely to send it to the help-seeker (M=3.67, SD=1.00), and thought it was the least problematic (M=2.01, SD=1.22) and least inappropriate (M=1.99, SD=1.18). Participants also liked the peer's response the least (M=3.08, SD=0.04), said they were least likely to send it (M=3.05, SD=1.15), and thought it was the least helpful (M=3.12, SD=1.09), the most problematic (M=2.25, SD=1.19) and the most inappropriate (M=2.26, SD=1.23).
See Figure \ref{fig:plot-self-expression} for a visual comparison of the means and standard deviations for reactions to all responses about self-expression. 

Participants liked that the \textbf{AI-generated response} conveyed guidance, support, and understanding. \textit{``I think a response like this can make the person feel like they've all been in this situation before, and there are points of empathy that they can draw on to guide them to be braver by drawing on the respondent's situation.''} (P43, Male). Similar to the message about relationships, participants liked that this response \textit{``[empathizes] with the friend while also providing practical solutions''} (P587, Female). 

They felt the response from the \textbf{adult mentor} also conveyed some affirmation of the help-seeker's struggles but to a lesser extent. They liked that \textit{``it starts with sympathy''} (P160, Male), but the overall tone felt \textit{``unattached.''} This distinction between sympathy and empathy is important to note - as sympathy is feeling for someone else while empathy is feeling with them. The adult mentor's response conveyed "sympathy," meaning they can see where the help-seeker is coming from, but \textit{not empathy}, which implies less personal experience with the matter at hand. Participants clearly communicated this sentiment by saying that \textit{``the victim is not understood''} (P383, Male) and that the message indicates \textit{``the person [who wrote it] has no idea of what the [help-seeker] is going through.''} (P611, Female).

Similar to the reactions about responses to the help-seeking message about relationships, participants suspected that the \textbf{therapist's} message came from a clinician, \textit{``The message looks like one a professional therapist would send...''} (P350, Male). As one participant said, \textit{``The response is appropriate because [it] is from [a] counselor, someone who has been able to tackle much of the issues by being exposed to different clients and knowing what to do for certain client till they reach a point of believing in themselves and having that courage to talk to anyone without any fear.''} (P403, Male).

The \textbf{peer's response} did not feel supportive enough to participants and did not do enough to calm the help-seeker, \textit{``I think that's too casual a response to really consider the issue from the perspective of the person pouring it out.''} (P21, Male). Moreover, they felt this response lacked depth and oversimplified the help-seeker's fear of self-expression. Like the adult mentor response, participants also felt there was not enough of an indication that the person who wrote the response truly understands what the help-seeker is going through, but less so because they can't empathize and more because they do not want to put effort into conveying that empathy, \textit{``It was too brief and nothing more of empathy felt in it. It was more like this person wanted to help but didn’t really put much 
effort...
[it had] a little spark of encouragement[,] but it could be better.''} (P124, Male).

\subsubsection{Physical Health}
\label{sec:physical-health}
\edit{How much participants liked each source's responses to the help-seeking message about physical health were significantly different (Bartlett's test: p < 0.05; \textit{F}(3.0,1376.3) = 71.06, p < 0.001) as were the extent to which participants found the responses helpful (Bartlett's test: p < 0.05; \textit{F}(3.0,1376.1) = 77.94, p < 0.001), problematic (Bartlett's test: p > 0.05; \textit{F}(3.0,2484.0) = 20.76, p < 0.001), inappropriate (Bartlett's test: p > 0.05; \textit{F}(3.0,2484) = 17.35, p < 0.001), and how likely they were to send the response (Bartlett's test: p < 0.05; \textit{F}(3.0,1375.8) = 63.92, p < 0.001).}

\begin{figure*}[hbt!]
    \centering
    \includegraphics[scale=0.65]{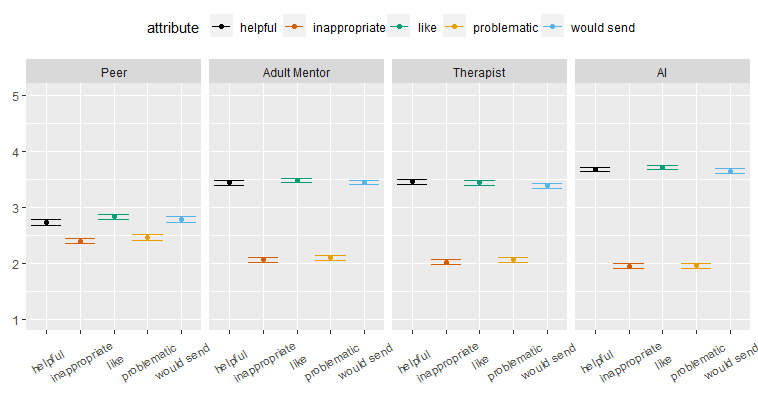}
    \caption{\edit{Participant preferences for human and AI-generated responses} by source for help-seeking message about physical health \edit{(RQ2, physical health)}. \edit{The y-axis is how much participants agreed with the different attributes of their reaction (i.e., how strongly they agreed with the statements described in section 3.2.2). Points represent means, and the bars on either end are standard deviation.}}
    \Description{The figure includes four dot plots with error bars for data from the scenario about physical health. Each plot corresponds to a different response source (peer, adult mentor, therapist, and AI). The x-axis lists different attributes of participants' reactions to help-giving responses (helpful, inappropriate, like, problematic, and would send) and the y-axis represents how much participants agreed with each attribute of their reaction (i.e., how strongly they agreed with the statements described in section 3.2.2).}
    \label{fig:plot-physical-health}
\end{figure*}

When the help-seeking message was about physical health, the AI responses once again outperformed the human-generated responses, especially the peer's. Participants liked AI's response the most (M=3.71, SD=0.96), saw it as the most helpful (M=3.67, SD=0.99), least problematic (M=1.95, SD=1.16), least inappropriate (M=1.95, SD=1.21), and overall felt they were most likely to send it to the help-seeker (M=3.64, SD=0.99). The peer response received the least favorable reaction as the least liked response (2.83, 0.05) and thought it was the most problematic (M=2.46, SD=1.28), most inappropriate (M=2.46, SD=1.28), and least helpful (M=2.73, SD=1.22), and indicated they were least likely to send it to the help-seeker (M=2.78, SD=1.23).
See Figure \ref{fig:plot-physical-health} for a visual comparison of the means and standard deviations for reactions to all responses about physical health.

The \textbf{AI-generated} response once again performed the best, and participants raved about how much they liked it - as one participant said, \textit{``I didn't just like this response. I love this response.''} (P351, Male).
Participants describe how the answer contained all the important components: empathy, acknowledgment, and concrete suggestions for next steps, \textit{``its an all round genuine answer [that] shows you care[,] understand the problem and offer the right solutions.''} (P592, Female). The "right solution," in this case, was to continue with ongoing self-care work (i.e., talking to a counselor), and participants liked that this response acknowledged doing so is not easy. \textit{``The fact that he empathized with the person and encouraged them to further try [counseling] because you cannot get the intended results after one trial.''} (P80, Male). Participants also liked that counseling was just one of the multiple solutions the response discussed and
\textit{``shows the [help-seeker] how they can use other strategies to be okay.''} (P453, Female). The comprehensive and deeply empathetic nature of this response seems to be what set it apart from the others.

Participants did like that the \textbf{adult mentor}'s response \textit{``[validated] wanting rest''} (P621, Non-binary), but also felt that it \textit{``[made] the [help-seeker] feel as if it's OK to just relax and not do anything.''} (P351, Male). They discussed how the adult mentor's message did not acknowledge that self-care is more than relaxation. In doing so, this response seemed to convey that there is nothing more the help-seeker can do, \textit{``I feel like it's not addressing the issue at hand and makes the victim feel like they are doing great of which they are not.''} (P137, Male). While participants thought \textit{``praising the [help-seeker] for small achievement[s]''} (P614, Female) and \textit{``positive energy''} were good, they were not enough, \textit{``because 
it doesn't contain enough [advice] that can help the victimized person''} (P125, Male).

While the \textbf{therapist}'s response did offer a solution, participants did not think it was the right one: \textit{``It's very unclear whether this respondent wants to help out. They are telling the victim to keep pursuing the same course that ain't working for them.''} (P167, Male). Unlike the AI-generated response, which suggested continued counseling in addition other solutions, participants did not like that the therapist's response was \textit{``just
telling the person in question to [continue]
counselling, [with] no other options to explore.''} (P122, Male). Therapy also incurs a cost for the help-seeker, which participants were not shy to mention in their feedback for this response. They noted that \textit{``counselling can be [quite] expensive,''} (P166, Male) and increasing those costs might not be necessary because \textit{``
at this point [the help-seeker] needs love and support from family and friends who would be available to support them.''} (P438, Female). This does not mean participants wanted the response to suggest the help-seeker stop counseling altogether, but rather that there should be additional non-clinical suggestions.



In summary, our findings revealed that participants preferred AI-generated responses \edit{to help-seeking messages about relationships, self-expression, and physical health}. However, \edit{when the help-seeking message was about suicidal thoughts}, participants preferred responses from the adult mentor. In the next section, we discuss the implications of these findings.
\section{Discussion}
Our findings point to several important areas for discussion regarding the promise of AI-powered support and questions to address about handling sensitive topics, should they arise.

\subsection{Premise for AI-Powered Support}

Our results provide insight about AI's potential to respond to help-seeking topics frequently represented in youth's conversations, which previous literature stated that they sometimes left without supportive responses. \edit{For instance, a study of private messages shared among youth online discovered norms of \textit{`unsupport'} when trust is hampered or when there is lack of resources or understanding of the situation by the youth peers \cite{huh2023help}. Rather than considering AI as a replacement for humans or directly giving support to the help-seeking individual, AI can play many roles in supporting the social support being exchanged among peers. Some examples include providing help-giving individuals tools to understand the intent of help-seeking individuals, ways to respond to the individuals in supportive ways, and how the responses can be stylistically modified to give better support. Our findings about participants' positive reactions to the AI-generated responses provide a premise for using generative AI to assist help-giving individuals in support-giving conversations, as long as the AI response or nudge can maintain} the characteristics that participants identified as helpful, empathetic, constructive, and non-judgemental. However, AI is a constantly moving target, where its characteristics, functionalities, and abilities constantly evolve over time. For instance, for Open AI's GPT models, the level of memorizing past conversations \edit{has been} limited, but recent updates enable\edit{d} the system to remember user preferences and tailor responses accordingly \cite{openai2023release}. Furthermore, depending on the version of GPT models, the prompt engineering techniques \edit{can generate different results}. Accordingly, the lesson we should take away from this study is to apply the rationale provided by the participants and devise ways that future generative AIs can maintain the preferred styles through customization, personalization, prompt engineering, \edit{using API wrappers to workaround shortcomings (e.g., \cite{trivedi2023openai}), and manipulation of training data. Doing so can help with improving the model's performances and transparency that leads to necessary levels of trust \cite{cabrero2023perceived} and human's reliance on AI systems \cite{kim2022perceived}}.

\edit{Some remaining questions include if participants knew it was AI, would they still be as positive as the results have shown? This question can be addressed from a future study, building on the efforts to improve AI-literacy \cite{lee2021developing} on ensuring there is no over- or under-trust of these systems. It seemed as though the participants felt genuine empathy from the AI responses, which was often missing in human-generated responses. If they find out that such genuine responses were artificially generated, it might or might not affect the trust in the system's responses. Our ongoing interview study with a subsample of these participants will address this issue.}




\subsection{AI as Complimentary Response based on the Context and Sensitivity of Topics}
Before considering the context of the help-seeking message, it would appear that young people prefer the adult mentor's response. However, a closer look reveals a strong preference for the adult mentor's response to the sensitive help-seeking message about suicidal thoughts while they preferred the AI-generated response to all other help-seeking messages. 
In a research article by Shao \cite{Shao2023empathetic}, a similar encounter takes place where their results indicate human-operated counselors provide higher quality results during empathetic counseling sessions than AI, but upon further investigation, it was clear that the way in which AI-supported an individual became crucial to its success in being rated helpful. 
As indicated by our findings and as previous research \cite{Saha2022towards,ahn2020chatbot} has uncovered participants usually preferred responses that offered support and comfort within the scope of positivity to feel comfortable enough to continue to share. 

We do not intend for our results to imply that AI-generated content is sufficient to replace a therapist or clinical expertise. As we found, participants did not like AI's response about sensitive topics,
for which ChatGPT's response specified that the kind of support required was outside the model's capabilities. While some participants appreciated the honest answer that reflected this limitation, others noted that referral to someone else felt dismissive because the help-seeker could have gone to that other source to begin with, but they did not. This points to two interesting follow-up questions: 1. if young people are experiencing suicidal thoughts, would they go to an AI system for help in a real-world situation? 2. if the answer to question 1 is yes, then how should an AI system respond when a user asks for help it cannot provide for sensitive matters? 
Researchers \cite{mahmood2022owning} have begun to explore how AI should own up to its faults during user interactions, such as when it may provide incorrect information. Our work highlights that similar frameworks should be explored for sensitive topics -- for example, is it more helpful for the AI to respond with links and contact information for resources? If so, which resources are most helpful to share and why? The goal is for AI to convey its limitations constructively, ensuring the user is safe.


The highly sensitive and complex issue of suicidal thoughts demands careful consideration of ethical, legal, and moral concerns, which the current AI technologies cannot address. In creating AI-based risk detection and prevention techniques, we need to keep humans in the loop, such as using AI as an assistant to peer support giver rather than directly having AI interact with struggling individuals as an all-in-one solution. Current AI technology may misinterpret language and is not able to understand the context of the situation, and nuances of human emotional state, which can lead to misidentification. Those false positives or negatives could have catastrophic results depending on the context. Having higher false alarms may burden resources depending on the types of interventions being implemented after detection, such as connecting them to mental health professionals. At the same time, missing instances may leave individuals in distress who may need help. 


Furthermore, even though our results indicate that young people prefer responses from an adult mentor, it is possible that participants did not like the idea that a potential peer would interact like a clinician. Participants often noted that the therapist's responses sounded "clinical," which is not a type of interaction that they would expect from peers. That violation of expectations could be what they responded negatively to. We want to stress that cases at risk of suicide require crisis intervention and professional solutions, such as therapy. AI conversational agents could be helpful in identifying users at risk of suicide and self-harm, but it is imperative that we not rely on AI alone to detect imminent risks. Furthermore, there are open questions about monitoring stigmatized topics like suicide, as it can perpetuate algorithmic stigmatization~\cite{Andalibi2023conceptualizing} for individuals struggling with mental health issues and create an environment where they feel constantly surveilled. \edit{Researchers should investigate the social implications of deploying AI in sensitive contexts such as mental health support based on four algorithmic stigma elements (i.e., labeling, stereotyping, separation, status loss/discrimination)~\cite{Andalibi2023conceptualizing} which include representational and/or allocative harms from the perspective of youth.}

\subsection{Design Implications}
Our findings show how AI-generated responses could be \edit{perceived as acceptable by young people. This finding provides implications for the possibility that AI-generated responses can be used to train or nudge youth to support their peers. AI could be specifically trained with best peer support practices, such as providing specialized recommendations for youth, following therapeutic approaches, or persuasive talks. For instance, behavior change literature has shown that young people value mental, social, financial, and spiritual aspects when conceptualizing health more broadly than physical health or prevention for chronic illness \cite{molenaar2020language}. Thus, persuasion of behavior change needs to be engaging, feasible, and acceptable \cite{molenaar2020language} rather than relying on the best practices for general populations.} 

\edit{Given the unique needs young people have about persuasion and conversations, rather than relying on expert- or other top-down approaches, humans can co-shape how AI can help in a personalized way. For instance, recent work has shown how humans can help facilitate AI's} expression of empathy in textual supportive conversations in noncritical settings~\cite{sharma2023human}. \edit{Numerous research studies have also explored efforts to generate creative content from human-AI collaborations \cite{hughes2021generative}. For instance, haiku poetry made with stimuli selected by humans were rated as more favorable than those that were generated solely by AI \cite{hitsuwari2023does}, and AI colleagues can help video game designers at various levels with design tasks \cite{lucas2017stay}. AI can provide resources and support suggestions, and young people can in return provide feedback about what has been helpful or not, further enabling themselves to take agency in training themselves to be better supporters} in their social networks. \edit{For such usage cases of having AI as an assistant for peer support givers, we should ensure the processes and suggestions are transparent and explainable~\cite{shneiderman2020bridging}. The AI assistant should provide where the suggestion is coming from, and what the reasoning is behind it, and nudge young people to educate themselves and also be reflective about any recommendations made. Furthermore, it would be critical to understand young people's acceptance of AI-generated recommendations and anchor their decisions.
A systematic review of 60 articles on the acceptance of AI showed that perceived usefulness, performance expectancy, attitudes, trust, and effort expectancy significantly and positively predicted behavioral intention, willingness, and use behaviour of AI across multiple industries \cite{kelly2023factors}. However, in some scenarios involving cultural factors, the need for human contact could not be replicated or replaced by AI, regardless of the perceived usefulness or perceived ease of use. One shortcoming of these findings is that they have not engaged younger populations. As such, more research is warranted to understand how such subpopulations might react to newer forms of AI assistance.}

\edit{Also, specific clinically oriented problems cannot be sufficiently addressed by AI or peer supporters, and AI should be able to appropriately acknowledge limitations and respond with valid resources on} diagnostic assessment, psychosocial evaluation, medication management, and therapy~\footnote{\url{https://edis.ifas.ufl.edu}}. 
Similar to how SAMHSA~\footnote{\url{https://www.samhsa.gov/brss-tacs/recovery-support-tools/peers}} defines effective peer support for trauma survivors for those in recovery, it is important that in the design of effective peer support interventions researchers consider the trauma-informed guidelines with youth~\cite{razi2023trauma}: providing a sense of safety; trustworthiness and transparency; collaboration and mutuality; empowerment, voice and choice; and awareness of cultural, historical, and gender issues.

\subsection{Limitations and Future Directions}
We recruited participants from our networks and snowball sampling. Therefore, we acknowledged that our dataset might not be a generalizable sample that accurately represents the youth population in the US. \edit{Although we did meet the minimum sample size for 95\% confidence based on our power analysis (see section 3.1), we did not recruit for a specific subject pool makeup by sex, gender, or location, so our demographics may not be representative of U.S. youth as a whole.} 
Also, we asked participants about static conversation responses, not continuous interactions, \finaledit{because we used a scenario-based design approach. Scenario-based design allows designers to ask stakeholders about multiple usage possibilities and elicit potential concerns. This allows researchers to trust the results within the scenarios presented while promoting raising usage questions for additional exploration beyond the given context \cite{rosson2009scenario}. As such, we recommend that }future research investigate youth preference in real-time conversations with additional levels of exchange. \finaledit{Additionally, these studies should include measures to assess the impact of extended human-AI interactions on psychological well-being}. \edit{Our survey design also used only one response from each source for each help-seeking message, and the next iteration of this study could explore multiple responses per source for each help-seeking message to account for variability.}
Although we included support-seeking topics from real youths' social media interactions\edit{, we only focused on prevalent topics and did not present less frequent ones,} so our study is limited to the topics we \edit{included}. Future research should study more comprehensive scenarios and go more in-depth on specific support-seeking topics to provide deeper insights into each topic. \edit{Future research could also address how} participants \edit{respond when they} kn\edit{o}w the origin of the responses and when they were interacting with AI. 
\edit{We did not collect qualitative data about every help-seeking message x response combination because survey completion rates drastically reduce as the total number of questions increases \cite{liu2018examining}.} To further understand youth's norms around peer support and perceptions of how AI could be helpful or inappropriate, we are planning to conduct interviews with some of the young people who participated in our study. \finaledit{Within our research setting, we could review AI-generated responses before including them in the survey to avoid presenting harmful material, but this level of human review is not commonplace in real-world human-AI interactions. As such, we implore researchers to explore potential limitations in AI systems' ability to understand and respond to complex human emotions to develop accountability measures for when AI might offer inappropriate or potentially harmful information.}

\section{Conclusion}
To explore the potential for AI-powered peer support, we studied how young people react to different human- and AI-generated responses to online help-seeking messages. Our findings revealed that participants preferred AI-generated responses about less sensitive topics (relationships, self-expression, and physical health), but not for sensitive discussions about suicidal thoughts. These findings point to the promise of AI-powered tools to help peers support each other during challenging times.
\begin{acks}
This research was supported in part by the National Science Foundation (award \#2144880). Any opinions, findings, conclusions, or recommendations expressed in this material are those of the authors and do not necessarily reflect the views of the research sponsors. Special thanks to all the participants who contributed their time and data.
\end{acks}
\bibliographystyle{ACM-Reference-Format}
\bibliography{references.bib,zotero_references.bib}


\begin{thebibliography}{91}


\ifx \showCODEN    \undefined \def \showCODEN     #1{\unskip}     \fi
\ifx \showDOI      \undefined \def \showDOI       #1{#1}\fi
\ifx \showISBNx    \undefined \def \showISBNx     #1{\unskip}     \fi
\ifx \showISBNxiii \undefined \def \showISBNxiii  #1{\unskip}     \fi
\ifx \showISSN     \undefined \def \showISSN      #1{\unskip}     \fi
\ifx \showLCCN     \undefined \def \showLCCN      #1{\unskip}     \fi
\ifx \shownote     \undefined \def \shownote      #1{#1}          \fi
\ifx \showarticletitle \undefined \def \showarticletitle #1{#1}   \fi
\ifx \showURL      \undefined \def \showURL       {\relax}        \fi
\providecommand\bibfield[2]{#2}
\providecommand\bibinfo[2]{#2}
\providecommand\natexlab[1]{#1}
\providecommand\showeprint[2][]{arXiv:#2}

\bibitem[sal(2023)]%
        {salesforce2023new}
 \bibinfo{year}{2023}\natexlab{}.
\newblock \bibinfo{title}{New AI Usage Data Shows Who’s Using AI — and
  Uncovers a Population of ‘Super-Users’}.
\newblock
\newblock
\urldef\tempurl%
\url{https://www.salesforce.com/news/press-releases/2023/09/07/ai-usage-research/}
\showURL{%
\tempurl}


\bibitem[Aggarwal et~al\mbox{.}(2023)]%
        {aggarwal2023artificial}
\bibfield{author}{\bibinfo{person}{Abhishek Aggarwal},
  \bibinfo{person}{Cheuk~Chi Tam}, \bibinfo{person}{Dezhi Wu},
  \bibinfo{person}{Xiaoming Li}, {and} \bibinfo{person}{Shan Qiao}.}
  \bibinfo{year}{2023}\natexlab{}.
\newblock \showarticletitle{Artificial Intelligence--Based Chatbots for
  Promoting Health Behavioral Changes: Systematic Review}.
\newblock \bibinfo{journal}{\emph{Journal of Medical Internet Research}}
  \bibinfo{volume}{25} (\bibinfo{year}{2023}), \bibinfo{pages}{e40789}.
\newblock


\bibitem[Ahmed et~al\mbox{.}(2022)]%
        {ahmed_thematic_2022}
\bibfield{author}{\bibinfo{person}{Arfan Ahmed}, \bibinfo{person}{Sarah Aziz},
  \bibinfo{person}{Mohamed Khalifa}, \bibinfo{person}{Uzair Shah},
  \bibinfo{person}{Asma Hassan}, \bibinfo{person}{Alaa Abd-Alrazaq}, {and}
  \bibinfo{person}{Mowafa Househ}.} \bibinfo{year}{2022}\natexlab{}.
\newblock \showarticletitle{Thematic {Analysis} on {User} {Reviews} for
  {Depression} and {Anxiety} {Chatbot} {Apps}: {Machine} {Learning}
  {Approach}}.
\newblock \bibinfo{journal}{\emph{JMIR Formative Research}}
  \bibinfo{volume}{6}, \bibinfo{number}{3} (\bibinfo{date}{March}
  \bibinfo{year}{2022}), \bibinfo{pages}{e27654}.
\newblock
\urldef\tempurl%
\url{https://doi.org/10.2196/27654}
\showDOI{\tempurl}
\newblock
\shownote{Company: JMIR Formative Research Distributor: JMIR Formative Research
  Institution: JMIR Formative Research Label: JMIR Formative Research
  Publisher: JMIR Publications Inc., Toronto, Canada}.


\bibitem[Ahn et~al\mbox{.}(2020)]%
        {ahn2020chatbot}
\bibfield{author}{\bibinfo{person}{Yuna Ahn}, \bibinfo{person}{Yilin Zhang},
  \bibinfo{person}{Yujin Park}, {and} \bibinfo{person}{Joonhwan Lee}.}
  \bibinfo{year}{2020}\natexlab{}.
\newblock \showarticletitle{A chatbot solution to chat app problems:
  Envisioning a chatbot counseling system for teenage victims of online sexual
  exploitation}. In \bibinfo{booktitle}{\emph{Extended Abstracts of the 2020
  CHI Conference on Human Factors in Computing Systems}}.
  \bibinfo{pages}{1--7}.
\newblock


\bibitem[Andalibi et~al\mbox{.}(2023)]%
        {Andalibi2023conceptualizing}
\bibfield{author}{\bibinfo{person}{Nazanin Andalibi}, \bibinfo{person}{Cassidy
  Pyle}, \bibinfo{person}{Kristen Barta}, \bibinfo{person}{Lu Xian},
  \bibinfo{person}{Abigail~Z. Jacobs}, {and} \bibinfo{person}{Mark~S.
  Ackerman}.} \bibinfo{year}{2023}\natexlab{}.
\newblock \showarticletitle{Conceptualizing Algorithmic Stigmatization}. In
  \bibinfo{booktitle}{\emph{Proceedings of the 2023 CHI Conference on Human
  Factors in Computing Systems}} (Hamburg, Germany) \emph{(\bibinfo{series}{CHI
  '23})}. \bibinfo{publisher}{Association for Computing Machinery},
  \bibinfo{address}{New York, NY, USA}, Article \bibinfo{articleno}{373},
  \bibinfo{numpages}{18}~pages.
\newblock
\showISBNx{9781450394215}
\urldef\tempurl%
\url{https://doi.org/10.1145/3544548.3580970}
\showDOI{\tempurl}


\bibitem[Armstrong and Powell(2009)]%
        {armstrong2009patient}
\bibfield{author}{\bibinfo{person}{Natalie Armstrong} {and}
  \bibinfo{person}{John Powell}.} \bibinfo{year}{2009}\natexlab{}.
\newblock \showarticletitle{Patient perspectives on health advice posted on
  Internet discussion boards: a qualitative study}.
\newblock \bibinfo{journal}{\emph{Health Expectations}} \bibinfo{volume}{12},
  \bibinfo{number}{3} (\bibinfo{year}{2009}), \bibinfo{pages}{313--320}.
\newblock


\bibitem[Auerbach et~al\mbox{.}(2018)]%
        {auerbach2018world}
\bibfield{author}{\bibinfo{person}{Randy~P Auerbach}, \bibinfo{person}{Philippe
  Mortier}, \bibinfo{person}{Ronny Bruffaerts}, \bibinfo{person}{Jordi Alonso},
  \bibinfo{person}{Corina Benjet}, \bibinfo{person}{Pim Cuijpers},
  \bibinfo{person}{Koen Demyttenaere}, \bibinfo{person}{David~D Ebert},
  \bibinfo{person}{Jennifer~Greif Green}, \bibinfo{person}{Penelope Hasking},
  {et~al\mbox{.}}} \bibinfo{year}{2018}\natexlab{}.
\newblock \showarticletitle{WHO world mental health surveys international
  college student project: prevalence and distribution of mental disorders.}
\newblock \bibinfo{journal}{\emph{Journal of abnormal psychology}}
  \bibinfo{volume}{127}, \bibinfo{number}{7} (\bibinfo{year}{2018}),
  \bibinfo{pages}{623}.
\newblock


\bibitem[Bae~Brandtz{\ae}g et~al\mbox{.}(2021)]%
        {bae2021social}
\bibfield{author}{\bibinfo{person}{Petter~Bae Bae~Brandtz{\ae}g},
  \bibinfo{person}{Marita Skjuve}, \bibinfo{person}{Kim~Kristoffer
  Kristoffer~Dysthe}, {and} \bibinfo{person}{Asbj{\o}rn F{\o}lstad}.}
  \bibinfo{year}{2021}\natexlab{}.
\newblock \showarticletitle{When the social becomes non-human: young people's
  perception of social support in chatbots}. In
  \bibinfo{booktitle}{\emph{Proceedings of the 2021 CHI conference on human
  factors in computing systems}}. \bibinfo{pages}{1--13}.
\newblock


\bibitem[Bechler et~al\mbox{.}(2021)]%
        {bechler2021attitude}
\bibfield{author}{\bibinfo{person}{Christopher~J Bechler},
  \bibinfo{person}{Zakary~L Tormala}, {and} \bibinfo{person}{Derek~D Rucker}.}
  \bibinfo{year}{2021}\natexlab{}.
\newblock \showarticletitle{The attitude--behavior relationship revisited}.
\newblock \bibinfo{journal}{\emph{Psychological science}} \bibinfo{volume}{32},
  \bibinfo{number}{8} (\bibinfo{year}{2021}), \bibinfo{pages}{1285--1297}.
\newblock


\bibitem[Bell et~al\mbox{.}(2018)]%
        {bell2018suicide}
\bibfield{author}{\bibinfo{person}{Jo Bell}, \bibinfo{person}{Katherine Mok},
  \bibinfo{person}{Eric Gardiner}, {and} \bibinfo{person}{Jane Pirkis}.}
  \bibinfo{year}{2018}\natexlab{}.
\newblock \showarticletitle{Suicide-related internet use among suicidal young
  people in the UK: Characteristics of users, effects of use, and barriers to
  offline help-seeking}.
\newblock \bibinfo{journal}{\emph{Archives of Suicide Research}}
  \bibinfo{volume}{22}, \bibinfo{number}{2} (\bibinfo{year}{2018}),
  \bibinfo{pages}{263--277}.
\newblock


\bibitem[Best et~al\mbox{.}(2016)]%
        {best2016seeking}
\bibfield{author}{\bibinfo{person}{Paul Best}, \bibinfo{person}{Elena
  Gil-Rodriguez}, \bibinfo{person}{Roger Manktelow}, {and}
  \bibinfo{person}{Brian~J Taylor}.} \bibinfo{year}{2016}\natexlab{}.
\newblock \showarticletitle{Seeking help from everyone and no-one:
  Conceptualizing the online help-seeking process among adolescent males}.
\newblock \bibinfo{journal}{\emph{Qualitative health research}}
  \bibinfo{volume}{26}, \bibinfo{number}{8} (\bibinfo{year}{2016}),
  \bibinfo{pages}{1067--1077}.
\newblock


\bibitem[Birnbaum et~al\mbox{.}(2017)]%
        {birnbaum2017role}
\bibfield{author}{\bibinfo{person}{Michael~L Birnbaum}, \bibinfo{person}{Asra~F
  Rizvi}, \bibinfo{person}{Christoph~U Correll}, \bibinfo{person}{John~M Kane},
  {and} \bibinfo{person}{Jamie Confino}.} \bibinfo{year}{2017}\natexlab{}.
\newblock \showarticletitle{Role of social media and the Internet in pathways
  to care for adolescents and young adults with psychotic disorders and
  non-psychotic mood disorders}.
\newblock \bibinfo{journal}{\emph{Early intervention in psychiatry}}
  \bibinfo{volume}{11}, \bibinfo{number}{4} (\bibinfo{year}{2017}),
  \bibinfo{pages}{290--295}.
\newblock


\bibitem[Bishop(2021)]%
        {bishop2021artificial}
\bibfield{author}{\bibinfo{person}{J~Mark Bishop}.}
  \bibinfo{year}{2021}\natexlab{}.
\newblock \showarticletitle{Artificial intelligence is stupid and causal
  reasoning will not fix it}.
\newblock \bibinfo{journal}{\emph{Frontiers in Psychology}}
  \bibinfo{volume}{11} (\bibinfo{year}{2021}), \bibinfo{pages}{2603}.
\newblock


\bibitem[Bitterly et~al\mbox{.}(2017)]%
        {bitterly2017risky}
\bibfield{author}{\bibinfo{person}{T~Bradford Bitterly},
  \bibinfo{person}{Alison~Wood Brooks}, {and} \bibinfo{person}{Maurice~E
  Schweitzer}.} \bibinfo{year}{2017}\natexlab{}.
\newblock \showarticletitle{Risky business: When humor increases and decreases
  status.}
\newblock \bibinfo{journal}{\emph{Journal of personality and social
  psychology}} \bibinfo{volume}{112}, \bibinfo{number}{3}
  (\bibinfo{year}{2017}), \bibinfo{pages}{431}.
\newblock


\bibitem[Bradford and Rickwood(2014)]%
        {bradford2014adolescent}
\bibfield{author}{\bibinfo{person}{Sally Bradford} {and} \bibinfo{person}{Debra
  Rickwood}.} \bibinfo{year}{2014}\natexlab{}.
\newblock \showarticletitle{Adolescent's preferred modes of delivery for mental
  health services}.
\newblock \bibinfo{journal}{\emph{Child and Adolescent Mental Health}}
  \bibinfo{volume}{19}, \bibinfo{number}{1} (\bibinfo{year}{2014}),
  \bibinfo{pages}{39--45}.
\newblock


\bibitem[Br{\aa}dvik(2018)]%
        {braadvik2018suicide}
\bibfield{author}{\bibinfo{person}{Louise Br{\aa}dvik}.}
  \bibinfo{year}{2018}\natexlab{}.
\newblock \bibinfo{title}{Suicide risk and mental disorders}.
\newblock , \bibinfo{numpages}{2028}~pages.
\newblock


\bibitem[Brandtzaeg and F{\o}lstad(2018)]%
        {brandtzaeg2018chatbots}
\bibfield{author}{\bibinfo{person}{Petter~Bae Brandtzaeg} {and}
  \bibinfo{person}{Asbj{\o}rn F{\o}lstad}.} \bibinfo{year}{2018}\natexlab{}.
\newblock \showarticletitle{Chatbots: changing user needs and motivations}.
\newblock \bibinfo{journal}{\emph{interactions}} \bibinfo{volume}{25},
  \bibinfo{number}{5} (\bibinfo{year}{2018}), \bibinfo{pages}{38--43}.
\newblock


\bibitem[Braun and Clarke(2019)]%
        {braun2019reflecting}
\bibfield{author}{\bibinfo{person}{Virginia Braun} {and}
  \bibinfo{person}{Victoria Clarke}.} \bibinfo{year}{2019}\natexlab{}.
\newblock \showarticletitle{Reflecting on reflexive thematic analysis}.
\newblock \bibinfo{journal}{\emph{Qualitative research in sport, exercise and
  health}} \bibinfo{volume}{11}, \bibinfo{number}{4} (\bibinfo{year}{2019}),
  \bibinfo{pages}{589--597}.
\newblock


\bibitem[Burns et~al\mbox{.}(2016)]%
        {burns2016role}
\bibfield{author}{\bibinfo{person}{Jane~M Burns}, \bibinfo{person}{Emma
  Birrell}, \bibinfo{person}{Marie Bismark}, \bibinfo{person}{Jane Pirkis},
  \bibinfo{person}{Tracey~A Davenport}, \bibinfo{person}{Ian~B Hickie},
  \bibinfo{person}{Melissa~K Weinberg}, {and} \bibinfo{person}{Louise~A
  Ellis}.} \bibinfo{year}{2016}\natexlab{}.
\newblock \showarticletitle{The role of technology in Australian youth mental
  health reform}.
\newblock \bibinfo{journal}{\emph{Australian Health Review}}
  \bibinfo{volume}{40}, \bibinfo{number}{5} (\bibinfo{year}{2016}),
  \bibinfo{pages}{584--590}.
\newblock


\bibitem[Burns et~al\mbox{.}(2010)]%
        {burns2010internet}
\bibfield{author}{\bibinfo{person}{Jane~M Burns}, \bibinfo{person}{Tracey~A
  Davenport}, \bibinfo{person}{Lauren~A Durkin}, \bibinfo{person}{Georgina~M
  Luscombe}, {and} \bibinfo{person}{Ian~B Hickie}.}
  \bibinfo{year}{2010}\natexlab{}.
\newblock \showarticletitle{The internet as a setting for mental health service
  utilisation by young people}.
\newblock \bibinfo{journal}{\emph{Medical Journal of Australia}}
  \bibinfo{volume}{192} (\bibinfo{year}{2010}), \bibinfo{pages}{S22--S26}.
\newblock


\bibitem[Cabrero-Daniel and Sanagust{\'\i}n~Cabrero(2023)]%
        {cabrero2023perceived}
\bibfield{author}{\bibinfo{person}{Beatriz Cabrero-Daniel} {and}
  \bibinfo{person}{Andrea Sanagust{\'\i}n~Cabrero}.}
  \bibinfo{year}{2023}\natexlab{}.
\newblock \showarticletitle{Perceived Trustworthiness of Natural Language
  Generators}. In \bibinfo{booktitle}{\emph{Proceedings of the First
  International Symposium on Trustworthy Autonomous Systems}}.
  \bibinfo{pages}{1--9}.
\newblock


\bibitem[Carlbring et~al\mbox{.}(2023)]%
        {carlbring2023new}
\bibfield{author}{\bibinfo{person}{Per Carlbring}, \bibinfo{person}{Heather
  Hadjistavropoulos}, \bibinfo{person}{Annet Kleiboer}, {and}
  \bibinfo{person}{Gerhard Andersson}.} \bibinfo{year}{2023}\natexlab{}.
\newblock \showarticletitle{A new era in Internet interventions: The advent of
  Chat-GPT and AI-assisted therapist guidance}.
\newblock \bibinfo{journal}{\emph{Internet Interventions}}
  \bibinfo{volume}{32} (\bibinfo{year}{2023}).
\newblock


\bibitem[Center({[n.\,d.]})]%
        {annie2023young}
\bibfield{author}{\bibinfo{person}{The Annie E. Casey Foundation Kids~Count
  Center}.} \bibinfo{year}{[n.\,d.]}\natexlab{}.
\newblock \bibinfo{title}{YOUNG ADULT POPULATION AGES 18 TO 24 BY RACE AND
  ETHNICITY IN UNITED STATES}.
\newblock
\newblock
\urldef\tempurl%
\url{https://datacenter.aecf.org/data/tables/11207-young-adult-population-ages-18-to-24-by-race-and-ethnicity#detailed/1/any/false/1095/7983/21595,21596}
\showURL{%
\tempurl}


\bibitem[Cunningham et~al\mbox{.}(2014)]%
        {cunningham2014internet}
\bibfield{author}{\bibinfo{person}{John~A Cunningham}, \bibinfo{person}{Amelia
  Gulliver}, \bibinfo{person}{Lou Farrer}, \bibinfo{person}{Kylie Bennett},
  {and} \bibinfo{person}{Bradley Carron-Arthur}.}
  \bibinfo{year}{2014}\natexlab{}.
\newblock \showarticletitle{Internet interventions for mental health and
  addictions: current findings and future directions}.
\newblock \bibinfo{journal}{\emph{Current psychiatry reports}}
  \bibinfo{volume}{16} (\bibinfo{year}{2014}), \bibinfo{pages}{1--5}.
\newblock


\bibitem[Curtin et~al\mbox{.}(2022)]%
        {curtin2022provisional}
\bibfield{author}{\bibinfo{person}{Sally Curtin}, \bibinfo{person}{Matthew
  Garnett}, {and} \bibinfo{person}{Farida Ahmad}.}
  \bibinfo{year}{2022}\natexlab{}.
\newblock \bibinfo{title}{Provisional Numbers and Rates of Suicide by Month and
  Demographic Characteristics: United States, 2021}.
\newblock
\newblock
\urldef\tempurl%
\url{https://www.cdc.gov/nchs/data/vsrr/vsrr024.pdf}
\showURL{%
\tempurl}


\bibitem[Dakof and Taylor(1990)]%
        {dakof1990victims}
\bibfield{author}{\bibinfo{person}{Gayle~A Dakof} {and}
  \bibinfo{person}{Shelley~E Taylor}.} \bibinfo{year}{1990}\natexlab{}.
\newblock \showarticletitle{Victims' perceptions of social support: What is
  helpful from whom?}
\newblock \bibinfo{journal}{\emph{Journal of personality and social
  psychology}} \bibinfo{volume}{58}, \bibinfo{number}{1}
  (\bibinfo{year}{1990}), \bibinfo{pages}{80}.
\newblock


\bibitem[De~Choudhury and De(2014)]%
        {de2014mental}
\bibfield{author}{\bibinfo{person}{Munmun De~Choudhury} {and}
  \bibinfo{person}{Sushovan De}.} \bibinfo{year}{2014}\natexlab{}.
\newblock \showarticletitle{Mental health discourse on reddit: Self-disclosure,
  social support, and anonymity}. In \bibinfo{booktitle}{\emph{Proceedings of
  the international AAAI conference on web and social media}},
  Vol.~\bibinfo{volume}{8}. \bibinfo{pages}{71--80}.
\newblock


\bibitem[Ellis et~al\mbox{.}(2013)]%
        {ellis2013young}
\bibfield{author}{\bibinfo{person}{Louise~A Ellis}, \bibinfo{person}{Philippa
  Collin}, \bibinfo{person}{Patrick~J Hurley}, \bibinfo{person}{Tracey~A
  Davenport}, \bibinfo{person}{Jane~M Burns}, {and} \bibinfo{person}{Ian~B
  Hickie}.} \bibinfo{year}{2013}\natexlab{}.
\newblock \showarticletitle{Young men’s attitudes and behaviour in relation
  to mental health and technology: implications for the development of online
  mental health services}.
\newblock \bibinfo{journal}{\emph{BMC psychiatry}} \bibinfo{volume}{13},
  \bibinfo{number}{1} (\bibinfo{year}{2013}), \bibinfo{pages}{1--10}.
\newblock


\bibitem[Eshghie and Eshghie(2023)]%
        {eshghie2023chatgpt}
\bibfield{author}{\bibinfo{person}{Mahshid Eshghie} {and}
  \bibinfo{person}{Mojtaba Eshghie}.} \bibinfo{year}{2023}\natexlab{}.
\newblock \showarticletitle{ChatGPT as a Therapist Assistant: A Suitability
  Study}.
\newblock \bibinfo{journal}{\emph{arXiv preprint arXiv:2304.09873}}
  (\bibinfo{year}{2023}).
\newblock


\bibitem[Fitzpatrick et~al\mbox{.}(2017)]%
        {fitzpatrick2017delivering}
\bibfield{author}{\bibinfo{person}{Kathleen~Kara Fitzpatrick},
  \bibinfo{person}{Alison Darcy}, {and} \bibinfo{person}{Molly Vierhile}.}
  \bibinfo{year}{2017}\natexlab{}.
\newblock \showarticletitle{Delivering cognitive behavior therapy to young
  adults with symptoms of depression and anxiety using a fully automated
  conversational agent (Woebot): a randomized controlled trial}.
\newblock \bibinfo{journal}{\emph{JMIR mental health}} \bibinfo{volume}{4},
  \bibinfo{number}{2} (\bibinfo{year}{2017}), \bibinfo{pages}{e7785}.
\newblock


\bibitem[F{\o}lstad and Brandtz{\ae}g(2017)]%
        {folstad2017chatbots}
\bibfield{author}{\bibinfo{person}{Asbj{\o}rn F{\o}lstad} {and}
  \bibinfo{person}{Petter~Bae Brandtz{\ae}g}.} \bibinfo{year}{2017}\natexlab{}.
\newblock \showarticletitle{Chatbots and the new world of HCI}.
\newblock \bibinfo{journal}{\emph{interactions}} \bibinfo{volume}{24},
  \bibinfo{number}{4} (\bibinfo{year}{2017}), \bibinfo{pages}{38--42}.
\newblock


\bibitem[Frazier et~al\mbox{.}(2016)]%
        {frazier2016internet}
\bibfield{author}{\bibinfo{person}{Patricia Frazier}, \bibinfo{person}{Derek
  Richards}, \bibinfo{person}{Jacqueline Mooney}, \bibinfo{person}{Stefan
  Hofmann}, \bibinfo{person}{Deborah Beidel}, {and} \bibinfo{person}{Patrick
  Palmieri}.} \bibinfo{year}{2016}\natexlab{}.
\newblock \showarticletitle{Internet-Delivered Treatment for Depression,
  Anxiety, and Stress in University Students: A Patient Preference Trial}.
\newblock \bibinfo{journal}{\emph{Iproceedings}}  \bibinfo{volume}{2}
  (\bibinfo{date}{12} \bibinfo{year}{2016}), \bibinfo{pages}{e5}.
\newblock
\urldef\tempurl%
\url{https://doi.org/10.2196/iproc.6090}
\showDOI{\tempurl}


\bibitem[Frost et~al\mbox{.}(2015)]%
        {frost2015self}
\bibfield{author}{\bibinfo{person}{Mareka Frost}, \bibinfo{person}{Leanne
  Casey}, {and} \bibinfo{person}{Natalie Rando}.}
  \bibinfo{year}{2015}\natexlab{}.
\newblock \showarticletitle{Self-injury, help-seeking, and the Internet}.
\newblock \bibinfo{journal}{\emph{Crisis}} (\bibinfo{year}{2015}).
\newblock


\bibitem[Fruitman(2023)]%
        {fruitman2023online}
\bibfield{author}{\bibinfo{person}{Kate Fruitman}.}
  \bibinfo{year}{2023}\natexlab{}.
\newblock \showarticletitle{Online Peer Support for Substance Use Disorders}.
\newblock In \bibinfo{booktitle}{\emph{Technology-Assisted Interventions for
  Substance Use Disorders}}. \bibinfo{publisher}{Springer},
  \bibinfo{pages}{31--39}.
\newblock


\bibitem[Fulmer(2019)]%
        {fulmer2019artificial}
\bibfield{author}{\bibinfo{person}{Russell Fulmer}.}
  \bibinfo{year}{2019}\natexlab{}.
\newblock \showarticletitle{Artificial intelligence and counseling: Four levels
  of implementation}.
\newblock \bibinfo{journal}{\emph{Theory \& Psychology}} \bibinfo{volume}{29},
  \bibinfo{number}{6} (\bibinfo{year}{2019}), \bibinfo{pages}{807--819}.
\newblock


\bibitem[Fulmer et~al\mbox{.}(2018)]%
        {fulmer2018using}
\bibfield{author}{\bibinfo{person}{Russell Fulmer}, \bibinfo{person}{Angela
  Joerin}, \bibinfo{person}{Breanna Gentile}, \bibinfo{person}{Lysanne
  Lakerink}, {and} \bibinfo{person}{Michiel Rauws}.}
  \bibinfo{year}{2018}\natexlab{}.
\newblock \showarticletitle{Using Psychological Artificial Intelligence (Tess)
  to Relieve Symptoms of Depression and Anxiety: Randomized Controlled Trial}.
\newblock \bibinfo{journal}{\emph{JMIR Ment Health}} \bibinfo{volume}{5},
  \bibinfo{number}{4} (\bibinfo{date}{13 Dec} \bibinfo{year}{2018}),
  \bibinfo{pages}{e64}.
\newblock
\showISSN{2368-7959}
\urldef\tempurl%
\url{https://doi.org/10.2196/mental.9782}
\showDOI{\tempurl}


\bibitem[G{\"o}tzl et~al\mbox{.}(2022)]%
        {gotzl2022artificial}
\bibfield{author}{\bibinfo{person}{Christian G{\"o}tzl},
  \bibinfo{person}{Selina Hiller}, \bibinfo{person}{Christian Rauschenberg},
  \bibinfo{person}{Anita Schick}, \bibinfo{person}{Janik Fechtelpeter},
  \bibinfo{person}{Unai Fischer~Abaigar}, \bibinfo{person}{Georgia Koppe},
  \bibinfo{person}{Daniel Durstewitz}, \bibinfo{person}{Ulrich Reininghaus},
  {and} \bibinfo{person}{Silvia Krumm}.} \bibinfo{year}{2022}\natexlab{}.
\newblock \showarticletitle{Artificial intelligence-informed mobile mental
  health apps for young people: a mixed-methods approach on users’ and
  stakeholders’ perspectives}.
\newblock \bibinfo{journal}{\emph{Child and Adolescent Psychiatry and Mental
  Health}} \bibinfo{volume}{16}, \bibinfo{number}{1} (\bibinfo{year}{2022}),
  \bibinfo{pages}{1--19}.
\newblock


\bibitem[Gould et~al\mbox{.}(2022)]%
        {gould2022crisis}
\bibfield{author}{\bibinfo{person}{Madelyn~S Gould}, \bibinfo{person}{Anthony
  Pisani}, \bibinfo{person}{Carlos Gallo}, \bibinfo{person}{Ashkan Ertefaie},
  \bibinfo{person}{Donald Harrington}, \bibinfo{person}{Caroline Kelberman},
  {and} \bibinfo{person}{Shannon Green}.} \bibinfo{year}{2022}\natexlab{}.
\newblock \showarticletitle{Crisis text-line interventions: Evaluation of
  texters' perceptions of effectiveness}.
\newblock \bibinfo{journal}{\emph{Suicide and Life-Threatening Behavior}}
  \bibinfo{volume}{52}, \bibinfo{number}{3} (\bibinfo{year}{2022}),
  \bibinfo{pages}{583--595}.
\newblock


\bibitem[Haque and Rubya(2023)]%
        {haque2023overview}
\bibfield{author}{\bibinfo{person}{MD~Romael Haque} {and}
  \bibinfo{person}{Sabirat Rubya}.} \bibinfo{year}{2023}\natexlab{}.
\newblock \showarticletitle{An Overview of Chatbot-Based Mobile Mental Health
  Apps: Insights From App Description and User Reviews}.
\newblock \bibinfo{journal}{\emph{JMIR mHealth and uHealth}}
  \bibinfo{volume}{11}, \bibinfo{number}{1} (\bibinfo{year}{2023}),
  \bibinfo{pages}{e44838}.
\newblock


\bibitem[Hartikainen et~al\mbox{.}(2021)]%
        {Hartikainen2021safesexting}
\bibfield{author}{\bibinfo{person}{Heidi Hartikainen}, \bibinfo{person}{Afsaneh
  Razi}, {and} \bibinfo{person}{Pamela Wisniewski}.}
  \bibinfo{year}{2021}\natexlab{}.
\newblock \showarticletitle{Safe Sexting: The Advice and Support Adolescents
  Receive from Peers Regarding Online Sexual Risks}.
\newblock \bibinfo{journal}{\emph{Proc. ACM Hum.-Comput. Interact.}}
  \bibinfo{volume}{5}, \bibinfo{number}{CSCW1}, Article \bibinfo{articleno}{42}
  (\bibinfo{date}{apr} \bibinfo{year}{2021}), \bibinfo{numpages}{31}~pages.
\newblock
\urldef\tempurl%
\url{https://doi.org/10.1145/3449116}
\showDOI{\tempurl}


\bibitem[Hitsuwari et~al\mbox{.}(2023)]%
        {hitsuwari2023does}
\bibfield{author}{\bibinfo{person}{Jimpei Hitsuwari},
  \bibinfo{person}{Yoshiyuki Ueda}, \bibinfo{person}{Woojin Yun}, {and}
  \bibinfo{person}{Michio Nomura}.} \bibinfo{year}{2023}\natexlab{}.
\newblock \showarticletitle{Does human--AI collaboration lead to more creative
  art? Aesthetic evaluation of human-made and AI-generated haiku poetry}.
\newblock \bibinfo{journal}{\emph{Computers in Human Behavior}}
  \bibinfo{volume}{139} (\bibinfo{year}{2023}), \bibinfo{pages}{107502}.
\newblock


\bibitem[Horgan and Sweeney(2010)]%
        {horgan2010young}
\bibfield{author}{\bibinfo{person}{Aine Horgan} {and} \bibinfo{person}{John
  Sweeney}.} \bibinfo{year}{2010}\natexlab{}.
\newblock \showarticletitle{Young students' use of the Internet for mental
  health information and support}.
\newblock \bibinfo{journal}{\emph{Journal of psychiatric and mental health
  nursing}} \bibinfo{volume}{17}, \bibinfo{number}{2} (\bibinfo{year}{2010}),
  \bibinfo{pages}{117--123}.
\newblock


\bibitem[Hughes et~al\mbox{.}(2021)]%
        {hughes2021generative}
\bibfield{author}{\bibinfo{person}{Rowan~T Hughes}, \bibinfo{person}{Liming
  Zhu}, {and} \bibinfo{person}{Tomasz Bednarz}.}
  \bibinfo{year}{2021}\natexlab{}.
\newblock \showarticletitle{Generative adversarial networks--enabled
  human--artificial intelligence collaborative applications for creative and
  design industries: A systematic review of current approaches and trends}.
\newblock \bibinfo{journal}{\emph{Frontiers in artificial intelligence}}
  \bibinfo{volume}{4} (\bibinfo{year}{2021}), \bibinfo{pages}{604234}.
\newblock


\bibitem[Huh-Yoo et~al\mbox{.}(2023)]%
        {huh2023help}
\bibfield{author}{\bibinfo{person}{Jina Huh-Yoo}, \bibinfo{person}{Afsaneh
  Razi}, \bibinfo{person}{Diep~N Nguyen}, \bibinfo{person}{Sampada Regmi},
  {and} \bibinfo{person}{Pamela~J Wisniewski}.}
  \bibinfo{year}{2023}\natexlab{}.
\newblock \showarticletitle{“Help Me:” Examining Youth’s Private Pleas
  for Support and the Responses Received from Peers via Instagram Direct
  Messages}. In \bibinfo{booktitle}{\emph{Proceedings of the 2023 CHI
  Conference on Human Factors in Computing Systems}}. \bibinfo{pages}{1--14}.
\newblock


\bibitem[Imran et~al\mbox{.}(2023)]%
        {imran2023chat}
\bibfield{author}{\bibinfo{person}{Nazish Imran}, \bibinfo{person}{Aateqa
  Hashmi}, {and} \bibinfo{person}{Ahad Imran}.}
  \bibinfo{year}{2023}\natexlab{}.
\newblock \showarticletitle{Chat-GPT: Opportunities and Challenges in Child
  Mental Healthcare}.
\newblock \bibinfo{journal}{\emph{Pakistan Journal of Medical Sciences}}
  \bibinfo{volume}{39}, \bibinfo{number}{4} (\bibinfo{year}{2023}).
\newblock


\bibitem[Jurewicz(2015)]%
        {jurewicz2015mental}
\bibfield{author}{\bibinfo{person}{Izabela Jurewicz}.}
  \bibinfo{year}{2015}\natexlab{}.
\newblock \showarticletitle{Mental health in young adults and
  adolescents--supporting general physicians to provide holistic care}.
\newblock \bibinfo{journal}{\emph{Clinical Medicine}} \bibinfo{volume}{15},
  \bibinfo{number}{2} (\bibinfo{year}{2015}), \bibinfo{pages}{151}.
\newblock


\bibitem[Kanagaraj(2020)]%
        {kanagaraj2020why}
\bibfield{author}{\bibinfo{person}{Manoj Kanagaraj}.}
  \bibinfo{year}{2020}\natexlab{}.
\newblock \bibinfo{title}{Here’s Why Mental Healthcare Is So Unaffordable and
  How COVID-19 Might Help Change This}.
\newblock
\newblock
\urldef\tempurl%
\url{https://info.primarycare.hms.harvard.edu/review/mental-health-unaffordable}
\showURL{%
\tempurl}


\bibitem[Kelly et~al\mbox{.}(2023)]%
        {kelly2023factors}
\bibfield{author}{\bibinfo{person}{Sage Kelly}, \bibinfo{person}{Sherrie-Anne
  Kaye}, {and} \bibinfo{person}{Oscar Oviedo-Trespalacios}.}
  \bibinfo{year}{2023}\natexlab{}.
\newblock \showarticletitle{What factors contribute to the acceptance of
  artificial intelligence? A systematic review}.
\newblock \bibinfo{journal}{\emph{Telematics and Informatics}}
  \bibinfo{volume}{77} (\bibinfo{year}{2023}), \bibinfo{pages}{101925}.
\newblock


\bibitem[Kemp and Henderson(2012)]%
        {kemp2012}
\bibfield{author}{\bibinfo{person}{Vivien Kemp} {and} \bibinfo{person}{Anthony
  Henderson}.} \bibinfo{year}{2012}\natexlab{}.
\newblock \showarticletitle{Challenges Faced by Mental Health Peer Support
  Workers: Peer Support from the Peer Supporter's Point of View}.
\newblock \bibinfo{journal}{\emph{Psychiatric rehabilitation journal}}
  \bibinfo{volume}{35} (\bibinfo{date}{03} \bibinfo{year}{2012}),
  \bibinfo{pages}{337--40}.
\newblock
\urldef\tempurl%
\url{https://doi.org/10.2975/35.4.2012.337.340}
\showDOI{\tempurl}


\bibitem[Khandpur(2017)]%
        {khandpur2017telemedicine}
\bibfield{author}{\bibinfo{person}{Raghbir~Singh Khandpur}.}
  \bibinfo{year}{2017}\natexlab{}.
\newblock \bibinfo{booktitle}{\emph{Telemedicine technology and applications
  (mHealth, TeleHealth and eHealth)}}.
\newblock \bibinfo{publisher}{PHI Learning Pvt. Ltd.}
\newblock


\bibitem[Kim et~al\mbox{.}(2022)]%
        {kim2022perceived}
\bibfield{author}{\bibinfo{person}{Jihyun Kim}, \bibinfo{person}{Kelly
  Merrill~Jr.}, \bibinfo{person}{Kun Xu}, {and} \bibinfo{person}{Stephanie
  Kelly}.} \bibinfo{year}{2022}\natexlab{}.
\newblock \showarticletitle{Perceived Credibility of an AI Instructor in Online
  Education: The Role of Social Presence and Voice Features}.
\newblock \bibinfo{journal}{\emph{Comput. Hum. Behav.}} \bibinfo{volume}{136},
  \bibinfo{number}{C} (\bibinfo{date}{nov} \bibinfo{year}{2022}),
  \bibinfo{numpages}{9}~pages.
\newblock
\showISSN{0747-5632}
\urldef\tempurl%
\url{https://doi.org/10.1016/j.chb.2022.107383}
\showDOI{\tempurl}


\bibitem[Kizito(2023)]%
        {kizito2023chatgpt}
\bibfield{author}{\bibinfo{person}{Samuel Kizito}.}
  \bibinfo{year}{2023}\natexlab{}.
\newblock \showarticletitle{ChatGPT has the potential to enhance antiretroviral
  therapy adherence among adolescents with HIV in sub-Saharan Africa}.
\newblock \bibinfo{journal}{\emph{Medical Education Online}}
  \bibinfo{volume}{28}, \bibinfo{number}{1} (\bibinfo{year}{2023}),
  \bibinfo{pages}{2246781}.
\newblock


\bibitem[Koulouri et~al\mbox{.}(2022)]%
        {koulouri2022chatbots}
\bibfield{author}{\bibinfo{person}{Theodora Koulouri},
  \bibinfo{person}{Robert~D Macredie}, {and} \bibinfo{person}{David Olakitan}.}
  \bibinfo{year}{2022}\natexlab{}.
\newblock \showarticletitle{Chatbots to support young adults’ mental health:
  an exploratory study of acceptability}.
\newblock \bibinfo{journal}{\emph{ACM Transactions on Interactive Intelligent
  Systems (TiiS)}} \bibinfo{volume}{12}, \bibinfo{number}{2}
  (\bibinfo{year}{2022}), \bibinfo{pages}{1--39}.
\newblock


\bibitem[Kuru and Pasek(2016)]%
        {kuru2016improving}
\bibfield{author}{\bibinfo{person}{Ozan Kuru} {and} \bibinfo{person}{Josh
  Pasek}.} \bibinfo{year}{2016}\natexlab{}.
\newblock \showarticletitle{Improving social media measurement in surveys:
  Avoiding acquiescence bias in Facebook research}.
\newblock \bibinfo{journal}{\emph{Computers in Human Behavior}}
  \bibinfo{volume}{57} (\bibinfo{year}{2016}), \bibinfo{pages}{82--92}.
\newblock


\bibitem[Lee et~al\mbox{.}(2021)]%
        {lee2021developing}
\bibfield{author}{\bibinfo{person}{Irene Lee}, \bibinfo{person}{Safinah Ali},
  \bibinfo{person}{Helen Zhang}, \bibinfo{person}{Daniella DiPaola}, {and}
  \bibinfo{person}{Cynthia Breazeal}.} \bibinfo{year}{2021}\natexlab{}.
\newblock \showarticletitle{Developing middle school students' AI literacy}. In
  \bibinfo{booktitle}{\emph{Proceedings of the 52nd ACM technical symposium on
  computer science education}}. \bibinfo{pages}{191--197}.
\newblock


\bibitem[Line({[n.\,d.]})]%
        {crisis2023decade}
\bibfield{author}{\bibinfo{person}{Crisis~Text Line}.}
  \bibinfo{year}{[n.\,d.]}\natexlab{}.
\newblock \bibinfo{title}{A Decade of Impact: Crisis Text Line’s 10-Year
  Impact Report}.
\newblock
\newblock
\urldef\tempurl%
\url{https://www.crisistextline.org/wp-content/uploads/2023/10/A-Decade-of-Impact-Report.pdf}
\showURL{%
\tempurl}


\bibitem[Liu and Wronski(2018)]%
        {liu2018examining}
\bibfield{author}{\bibinfo{person}{Mingnan Liu} {and} \bibinfo{person}{Laura
  Wronski}.} \bibinfo{year}{2018}\natexlab{}.
\newblock \showarticletitle{Examining completion rates in web surveys via over
  25,000 real-world surveys}.
\newblock \bibinfo{journal}{\emph{Social Science Computer Review}}
  \bibinfo{volume}{36}, \bibinfo{number}{1} (\bibinfo{year}{2018}),
  \bibinfo{pages}{116--124}.
\newblock


\bibitem[Lucas and Martinho(2017)]%
        {lucas2017stay}
\bibfield{author}{\bibinfo{person}{Pedro Lucas} {and} \bibinfo{person}{Carlos
  Martinho}.} \bibinfo{year}{2017}\natexlab{}.
\newblock \showarticletitle{Stay Awhile and Listen to 3Buddy, a Co-creative
  Level Design Support Tool.}. In \bibinfo{booktitle}{\emph{ICCC}}.
  \bibinfo{pages}{205--212}.
\newblock


\bibitem[Mahmood et~al\mbox{.}(2022)]%
        {mahmood2022owning}
\bibfield{author}{\bibinfo{person}{Amama Mahmood}, \bibinfo{person}{Jeanie~W
  Fung}, \bibinfo{person}{Isabel Won}, {and} \bibinfo{person}{Chien-Ming
  Huang}.} \bibinfo{year}{2022}\natexlab{}.
\newblock \showarticletitle{Owning mistakes sincerely: Strategies for
  mitigating AI errors}. In \bibinfo{booktitle}{\emph{Proceedings of the 2022
  CHI Conference on Human Factors in Computing Systems}}.
  \bibinfo{pages}{1--11}.
\newblock


\bibitem[Mar et~al\mbox{.}(2014)]%
        {mar2014exploring}
\bibfield{author}{\bibinfo{person}{Marissa~Y Mar}, \bibinfo{person}{Erika~K
  Neilson}, \bibinfo{person}{Iris Torchalla}, \bibinfo{person}{Gregory~R
  Werker}, \bibinfo{person}{Allison Laing}, {and} \bibinfo{person}{Michael
  Krausz}.} \bibinfo{year}{2014}\natexlab{}.
\newblock \showarticletitle{Exploring e-mental health preferences of generation
  Y}.
\newblock \bibinfo{journal}{\emph{Journal of Technology in Human Services}}
  \bibinfo{volume}{32}, \bibinfo{number}{4} (\bibinfo{year}{2014}),
  \bibinfo{pages}{312--327}.
\newblock


\bibitem[Mardikaningsih et~al\mbox{.}(2023)]%
        {mardikaningsih2023risk}
\bibfield{author}{\bibinfo{person}{Agusti Mardikaningsih},
  \bibinfo{person}{Ambiro~Puji Asmaroini}, \bibinfo{person}{Ahmad Lani}, {and}
  \bibinfo{person}{Raup Padillah}.} \bibinfo{year}{2023}\natexlab{}.
\newblock \showarticletitle{The risk of ChatGPT as a first aid method:
  considering the potential danger for mental health support among young adults
  without professional assistance}.
\newblock \bibinfo{journal}{\emph{Journal of Public Health}}
  (\bibinfo{year}{2023}), \bibinfo{pages}{fdad107}.
\newblock


\bibitem[Martinengo et~al\mbox{.}(2022)]%
        {martinengo2022evaluation}
\bibfield{author}{\bibinfo{person}{Laura Martinengo}, \bibinfo{person}{Elaine
  Lum}, {and} \bibinfo{person}{Josip Car}.} \bibinfo{year}{2022}\natexlab{}.
\newblock \showarticletitle{Evaluation of chatbot-delivered interventions for
  self-management of depression: content analysis}.
\newblock \bibinfo{journal}{\emph{Journal of affective disorders}}
  \bibinfo{volume}{319} (\bibinfo{year}{2022}), \bibinfo{pages}{598--607}.
\newblock


\bibitem[Molenaar et~al\mbox{.}(2020)]%
        {molenaar2020language}
\bibfield{author}{\bibinfo{person}{Annika Molenaar}, \bibinfo{person}{Tammie~ST
  Choi}, \bibinfo{person}{Linda Brennan}, \bibinfo{person}{Mike Reid},
  \bibinfo{person}{Megan~SC Lim}, \bibinfo{person}{Helen Truby}, {and}
  \bibinfo{person}{Tracy~A McCaffrey}.} \bibinfo{year}{2020}\natexlab{}.
\newblock \showarticletitle{Language of health of young Australian adults: a
  qualitative exploration of perceptions of health, wellbeing and health
  promotion via online conversations}.
\newblock \bibinfo{journal}{\emph{Nutrients}} \bibinfo{volume}{12},
  \bibinfo{number}{4} (\bibinfo{year}{2020}), \bibinfo{pages}{887}.
\newblock


\bibitem[Naslund et~al\mbox{.}(2016)]%
        {naslund2016future}
\bibfield{author}{\bibinfo{person}{John~A Naslund}, \bibinfo{person}{Kelly~A
  Aschbrenner}, \bibinfo{person}{Lisa~A Marsch}, {and}
  \bibinfo{person}{Stephen~J Bartels}.} \bibinfo{year}{2016}\natexlab{}.
\newblock \showarticletitle{The future of mental health care: peer-to-peer
  support and social media}.
\newblock \bibinfo{journal}{\emph{Epidemiology and psychiatric sciences}}
  \bibinfo{volume}{25}, \bibinfo{number}{2} (\bibinfo{year}{2016}),
  \bibinfo{pages}{113--122}.
\newblock


\bibitem[Natalie({[n.\,d.]})]%
        {openai2023release}
\bibfield{author}{\bibinfo{person}{Natalie}.}
  \bibinfo{year}{[n.\,d.]}\natexlab{}.
\newblock \bibinfo{title}{ChatGPT - Release Notes}.
\newblock
\newblock
\urldef\tempurl%
\url{https://help.openai.com/en/articles/6825453-chatgpt-release-notes}
\showURL{%
\tempurl}


\bibitem[Orenstein and Lewis(2022)]%
        {orenstein_eriksons_2022}
\bibfield{author}{\bibinfo{person}{Gabriel~A. Orenstein} {and}
  \bibinfo{person}{Lindsay Lewis}.} \bibinfo{year}{2022}\natexlab{}.
\newblock \showarticletitle{Eriksons {Stages} of {Psychosocial} {Development}}.
\newblock In \bibinfo{booktitle}{\emph{{StatPearls}}}.
  \bibinfo{publisher}{StatPearls Publishing}, \bibinfo{address}{Treasure Island
  (FL)}.
\newblock
\urldef\tempurl%
\url{http://www.ncbi.nlm.nih.gov/books/NBK556096/}
\showURL{%
\tempurl}


\bibitem[Pan and Zeng(2023)]%
        {pan2023llms}
\bibfield{author}{\bibinfo{person}{Keyu Pan} {and} \bibinfo{person}{Yawen
  Zeng}.} \bibinfo{year}{2023}\natexlab{}.
\newblock \showarticletitle{Do llms possess a personality? making the mbti test
  an amazing evaluation for large language models}.
\newblock \bibinfo{journal}{\emph{arXiv preprint arXiv:2307.16180}}
  (\bibinfo{year}{2023}).
\newblock


\bibitem[Pretorius et~al\mbox{.}(2019)]%
        {pretorius2019young}
\bibfield{author}{\bibinfo{person}{Claudette Pretorius}, \bibinfo{person}{Derek
  Chambers}, \bibinfo{person}{David Coyle}, {et~al\mbox{.}}}
  \bibinfo{year}{2019}\natexlab{}.
\newblock \showarticletitle{Young people’s online help-seeking and mental
  health difficulties: Systematic narrative review}.
\newblock \bibinfo{journal}{\emph{Journal of medical Internet research}}
  \bibinfo{volume}{21}, \bibinfo{number}{11} (\bibinfo{year}{2019}),
  \bibinfo{pages}{e13873}.
\newblock


\bibitem[Razi et~al\mbox{.}(2023)]%
        {razi2023aCSCW}
\bibfield{author}{\bibinfo{person}{Afsaneh Razi}, \bibinfo{person}{Ashwaq
  AlSoubai}, \bibinfo{person}{Seunghyun Kim}, \bibinfo{person}{Shiza Ali},
  \bibinfo{person}{Gianluca Stringhini}, \bibinfo{person}{Munmun Choudhury},
  {and} \bibinfo{person}{Pamela~J. Wisniewski}.}
  \bibinfo{year}{2023}\natexlab{}.
\newblock \showarticletitle{Sliding into My DMs: Detecting Uncomfortable or
  Unsafe Sexual Risk Experiences within Instagram Direct Messages Grounded in
  the Perspective of Youth}, Vol.~\bibinfo{volume}{7}.
  \bibinfo{publisher}{Association for Computing Machinery},
  \bibinfo{address}{New York, NY, USA}, Article \bibinfo{articleno}{89},
  \bibinfo{numpages}{28}~pages.
\newblock
\urldef\tempurl%
\url{https://doi.org/110.1145/3579522}
\showDOI{\tempurl}


\bibitem[Razi et~al\mbox{.}(2020)]%
        {razi_lets_2020}
\bibfield{author}{\bibinfo{person}{Afsaneh Razi}, \bibinfo{person}{Karla
  Badillo-Urquiola}, {and} \bibinfo{person}{Pamela~J. Wisniewski}.}
  \bibinfo{year}{2020}\natexlab{}.
\newblock \showarticletitle{Let's {Talk} about {Sext}: {How} {Adolescents}
  {Seek} {Support} and {Advice} about {Their} {Online} {Sexual} {Experiences}}.
  In \bibinfo{booktitle}{\emph{Proc 2020 {CHI} {Conference} on {Human}
  {Factors} in {Computing} {Systems}}} \emph{(\bibinfo{series}{{CHI} '20})}.
  \bibinfo{publisher}{ACM}, \bibinfo{address}{Honolulu, HI, USA},
  \bibinfo{pages}{1--13}.
\newblock
\showISBNx{978-1-4503-6708-0}
\urldef\tempurl%
\url{https://doi.org/10.1145/3313831.3376400}
\showDOI{\tempurl}


\bibitem[Razi et~al\mbox{.}(2021)]%
        {razi2021human}
\bibfield{author}{\bibinfo{person}{Afsaneh Razi}, \bibinfo{person}{Seunghyun
  Kim}, \bibinfo{person}{Ashwaq Alsoubai}, \bibinfo{person}{Gianluca
  Stringhini}, \bibinfo{person}{Thamar Solorio}, \bibinfo{person}{Munmun
  De~Choudhury}, {and} \bibinfo{person}{Pamela~J. Wisniewski}.}
  \bibinfo{year}{2021}\natexlab{}.
\newblock \showarticletitle{A Human-Centered Systematic Literature Review of
  the Computational Approaches for Online Sexual Risk Detection}.
\newblock \bibinfo{journal}{\emph{Proc. ACM Hum.-Comput. Interact.}}
  \bibinfo{volume}{5}, \bibinfo{number}{CSCW2}, Article
  \bibinfo{articleno}{465} (\bibinfo{date}{oct} \bibinfo{year}{2021}),
  \bibinfo{numpages}{38}~pages.
\newblock
\urldef\tempurl%
\url{https://doi.org/10.1145/3479609}
\showDOI{\tempurl}


\bibitem[Razi et~al\mbox{.}(2024)]%
        {razi2023trauma}
\bibfield{author}{\bibinfo{person}{Afsaneh Razi}, \bibinfo{person}{John
  Seberger}, \bibinfo{person}{Ashwaq Alsoubai}, \bibinfo{person}{Nurun Naher},
  \bibinfo{person}{Munmun De~Choudhury}, {and} \bibinfo{person}{Pamela~J.
  Wisniewski}.} \bibinfo{year}{2024}\natexlab{}.
\newblock \showarticletitle{Toward Trauma-Informed Research Practices with
  Youth in HCI: Caring for Participants and Research Assistants When Studying
  Sensitive Topics}.
\newblock \bibinfo{journal}{\emph{Proc. ACM Hum.-Comput. Interact.}}
  \bibinfo{volume}{8}, \bibinfo{number}{CSCW1}, Article
  \bibinfo{articleno}{134} (\bibinfo{year}{2024}),
  \bibinfo{numpages}{31}~pages.
\newblock
\urldef\tempurl%
\url{https://doi.org/10.1145/3637411}
\showURL{%
\tempurl}


\bibitem[Ridout et~al\mbox{.}(2018)]%
        {ridout2018use}
\bibfield{author}{\bibinfo{person}{Brad Ridout}, \bibinfo{person}{Andrew
  Campbell}, {et~al\mbox{.}}} \bibinfo{year}{2018}\natexlab{}.
\newblock \showarticletitle{The use of social networking sites in mental health
  interventions for young people: systematic review}.
\newblock \bibinfo{journal}{\emph{Journal of medical Internet research}}
  \bibinfo{volume}{20}, \bibinfo{number}{12} (\bibinfo{year}{2018}),
  \bibinfo{pages}{e12244}.
\newblock


\bibitem[Rita and Shava(2021)]%
        {rita2021chatbot}
\bibfield{author}{\bibinfo{person}{Marsela~Nur Rita} {and}
  \bibinfo{person}{Fungai~Bhunu Shava}.} \bibinfo{year}{2021}\natexlab{}.
\newblock \showarticletitle{Chatbot driven web-based platform for online safety
  and sexual exploitation awareness and reporting in Namibia}. In
  \bibinfo{booktitle}{\emph{2021 International Conference on Artificial
  Intelligence, Big Data, Computing and Data Communication Systems (icABCD)}}.
  IEEE, \bibinfo{pages}{1--5}.
\newblock


\bibitem[Rosson and Carroll(2009)]%
        {rosson2009scenario}
\bibfield{author}{\bibinfo{person}{Mary~Beth Rosson} {and}
  \bibinfo{person}{John~M Carroll}.} \bibinfo{year}{2009}\natexlab{}.
\newblock \showarticletitle{Scenario based design}.
\newblock \bibinfo{journal}{\emph{Human-computer interaction. boca raton, FL}}
  (\bibinfo{year}{2009}), \bibinfo{pages}{145--162}.
\newblock


\bibitem[Saha et~al\mbox{.}(2022)]%
        {Saha2022towards}
\bibfield{author}{\bibinfo{person}{Tulika Saha}, \bibinfo{person}{Vaibhav
  Gakhreja}, \bibinfo{person}{Anindya~Sundar Das}, \bibinfo{person}{Souhitya
  Chakraborty}, {and} \bibinfo{person}{Sriparna Saha}.}
  \bibinfo{year}{2022}\natexlab{}.
\newblock \showarticletitle{Towards Motivational and Empathetic Response
  Generation in Online Mental Health Support}. In
  \bibinfo{booktitle}{\emph{Proceedings of the 45th International ACM SIGIR
  Conference on Research and Development in Information Retrieval}} (Madrid,
  Spain) \emph{(\bibinfo{series}{SIGIR '22})}. \bibinfo{publisher}{Association
  for Computing Machinery}, \bibinfo{address}{New York, NY, USA},
  \bibinfo{pages}{2650–2656}.
\newblock
\showISBNx{9781450387323}
\urldef\tempurl%
\url{https://doi.org/10.1145/3477495.3531912}
\showDOI{\tempurl}


\bibitem[Saraceno(2002)]%
        {saraceno2002world}
\bibfield{author}{\bibinfo{person}{Benedetto Saraceno}.}
  \bibinfo{year}{2002}\natexlab{}.
\newblock \showarticletitle{The WHO world health report 2001 on mental health}.
\newblock \bibinfo{journal}{\emph{Epidemiology and Psychiatric Sciences}}
  \bibinfo{volume}{11}, \bibinfo{number}{2} (\bibinfo{year}{2002}),
  \bibinfo{pages}{83--87}.
\newblock


\bibitem[Sawyer et~al\mbox{.}(2018)]%
        {sawyer_age_2018}
\bibfield{author}{\bibinfo{person}{Susan~M. Sawyer}, \bibinfo{person}{Peter~S.
  Azzopardi}, \bibinfo{person}{Dakshitha Wickremarathne}, {and}
  \bibinfo{person}{George~C. Patton}.} \bibinfo{year}{2018}\natexlab{}.
\newblock \showarticletitle{The age of adolescence}.
\newblock \bibinfo{journal}{\emph{The Lancet Child \& Adolescent Health}}
  \bibinfo{volume}{2}, \bibinfo{number}{3} (\bibinfo{date}{March}
  \bibinfo{year}{2018}), \bibinfo{pages}{223--228}.
\newblock
\showISSN{2352-4642, 2352-4650}
\urldef\tempurl%
\url{https://doi.org/10.1016/S2352-4642(18)30022-1}
\showDOI{\tempurl}
\newblock
\shownote{Publisher: Elsevier}.


\bibitem[Shan et~al\mbox{.}(2022)]%
        {shan2022public}
\bibfield{author}{\bibinfo{person}{Yi Shan}, \bibinfo{person}{Meng Ji},
  \bibinfo{person}{Wenxiu Xie}, \bibinfo{person}{Kam-Yiu Lam}, {and}
  \bibinfo{person}{Chi-Yin Chow}.} \bibinfo{year}{2022}\natexlab{}.
\newblock \showarticletitle{Public trust in artificial intelligence
  applications in mental health care: topic modeling analysis}.
\newblock \bibinfo{journal}{\emph{JMIR Human Factors}} \bibinfo{volume}{9},
  \bibinfo{number}{4} (\bibinfo{year}{2022}), \bibinfo{pages}{e38799}.
\newblock


\bibitem[Shao(2023)]%
        {Shao2023empathetic}
\bibfield{author}{\bibinfo{person}{Ruosi Shao}.}
  \bibinfo{year}{2023}\natexlab{}.
\newblock \showarticletitle{An Empathetic AI for Mental Health Intervention:
  Conceptualizing and Examining Artificial Empathy}. In
  \bibinfo{booktitle}{\emph{Proceedings of the 2nd Empathy-Centric Design
  Workshop}} (Hamburg, Germany) \emph{(\bibinfo{series}{EMPATHICH '23})}.
  \bibinfo{publisher}{Association for Computing Machinery},
  \bibinfo{address}{New York, NY, USA}, Article \bibinfo{articleno}{4},
  \bibinfo{numpages}{6}~pages.
\newblock
\showISBNx{9798400707490}
\urldef\tempurl%
\url{https://doi.org/10.1145/3588967.3588971}
\showDOI{\tempurl}


\bibitem[Sharma et~al\mbox{.}(2023)]%
        {sharma2023human}
\bibfield{author}{\bibinfo{person}{Ashish Sharma}, \bibinfo{person}{Inna~W
  Lin}, \bibinfo{person}{Adam~S Miner}, \bibinfo{person}{David~C Atkins}, {and}
  \bibinfo{person}{Tim Althoff}.} \bibinfo{year}{2023}\natexlab{}.
\newblock \showarticletitle{Human--AI collaboration enables more empathic
  conversations in text-based peer-to-peer mental health support}.
\newblock \bibinfo{journal}{\emph{Nature Machine Intelligence}}
  \bibinfo{volume}{5}, \bibinfo{number}{1} (\bibinfo{year}{2023}),
  \bibinfo{pages}{46--57}.
\newblock


\bibitem[Shneiderman(2020)]%
        {shneiderman2020bridging}
\bibfield{author}{\bibinfo{person}{Ben Shneiderman}.}
  \bibinfo{year}{2020}\natexlab{}.
\newblock \showarticletitle{Bridging the gap between ethics and practice:
  guidelines for reliable, safe, and trustworthy human-centered AI systems}.
\newblock \bibinfo{journal}{\emph{ACM Transactions on Interactive Intelligent
  Systems (TiiS)}} \bibinfo{volume}{10}, \bibinfo{number}{4}
  (\bibinfo{year}{2020}), \bibinfo{pages}{1--31}.
\newblock


\bibitem[Singh(2023)]%
        {singh2023artificial}
\bibfield{author}{\bibinfo{person}{Om~P Singh}.}
  \bibinfo{year}{2023}\natexlab{}.
\newblock \showarticletitle{Artificial intelligence in the era of
  ChatGPT-Opportunities and challenges in mental health care}.
\newblock \bibinfo{journal}{\emph{Indian Journal of Psychiatry}}
  \bibinfo{volume}{65}, \bibinfo{number}{3} (\bibinfo{year}{2023}),
  \bibinfo{pages}{297}.
\newblock


\bibitem[{Statista}(2021)]%
        {statista2021}
\bibfield{author}{\bibinfo{person}{{Statista}}.}
  \bibinfo{year}{2021}\natexlab{}.
\newblock \bibinfo{booktitle}{\emph{Worldwide visits to Reddit.com from
  November 2022 to April 2023}}.
\newblock
\urldef\tempurl%
\url{https://www.statista.com/statistics/443332/reddit-monthly-visitors/}
\showURL{%
\tempurl}
\newblock
\shownote{Accessed on September 14, 2023}.


\bibitem[Trivedi et~al\mbox{.}(2023)]%
        {trivedi2023openai}
\bibfield{author}{\bibinfo{person}{Nilesh Trivedi}, \bibinfo{person}{Erik
  Berlin}, {and} \bibinfo{person}{Tom Lehman}.}
  \bibinfo{year}{2023}\natexlab{}.
\newblock \bibinfo{title}{OpenAI API client library to access GPT-3 in Ruby}.
\newblock
\newblock
\urldef\tempurl%
\url{https://github.com/nileshtrivedi/openai}
\showURL{%
\tempurl}


\bibitem[Uehara(1990)]%
        {uehara1990dual}
\bibfield{author}{\bibinfo{person}{Edwina Uehara}.}
  \bibinfo{year}{1990}\natexlab{}.
\newblock \showarticletitle{Dual exchange theory, social networks, and informal
  social support}.
\newblock \bibinfo{journal}{\emph{Amer. J. Sociology}} \bibinfo{volume}{96},
  \bibinfo{number}{3} (\bibinfo{year}{1990}), \bibinfo{pages}{521--557}.
\newblock


\bibitem[V{\"a}{\"a}n{\"a}nen et~al\mbox{.}(2020)]%
        {vaananen2020civicbots}
\bibfield{author}{\bibinfo{person}{Kaisa V{\"a}{\"a}n{\"a}nen},
  \bibinfo{person}{Aleksi Hiltunen}, \bibinfo{person}{Jari Varsaluoma}, {and}
  \bibinfo{person}{Iikka Pietil{\"a}}.} \bibinfo{year}{2020}\natexlab{}.
\newblock \showarticletitle{CivicBots--Chatbots for supporting youth in
  societal participation}. In \bibinfo{booktitle}{\emph{Chatbot Research and
  Design: Third International Workshop, CONVERSATIONS 2019, Amsterdam, The
  Netherlands, November 19--20, 2019, Revised Selected Papers 3}}. Springer,
  \bibinfo{pages}{143--157}.
\newblock


\bibitem[van Schalkwyk(2023)]%
        {van2023artificial}
\bibfield{author}{\bibinfo{person}{Gerrit van Schalkwyk}.}
  \bibinfo{year}{2023}\natexlab{}.
\newblock \bibinfo{title}{Artificial intelligence in pediatric behavioral
  health}.
\newblock , \bibinfo{numpages}{2}~pages.
\newblock


\bibitem[Vogels(2023)]%
        {vogels2023majority}
\bibfield{author}{\bibinfo{person}{Emily Vogels}.}
  \bibinfo{year}{2023}\natexlab{}.
\newblock \bibinfo{title}{A majority of Americans have heard of ChatGPT, but
  few have tried it themselves}.
\newblock
\newblock
\urldef\tempurl%
\url{https://www.pewresearch.org/short-reads/2023/05/24/a-majority-of-americans-have-heard-of-chatgpt-but-few-have-tried-it-themselves/}
\showURL{%
\tempurl}


\bibitem[Vogels et~al\mbox{.}(2022)]%
        {vogels2022teens}
\bibfield{author}{\bibinfo{person}{Emily~A Vogels}, \bibinfo{person}{Risa
  Gelles-Watnick}, {and} \bibinfo{person}{Navid Massarat}.}
  \bibinfo{year}{2022}\natexlab{}.
\newblock \showarticletitle{Teens, social media and technology 2022}.
\newblock  (\bibinfo{year}{2022}).
\newblock


\bibitem[Woodruff et~al\mbox{.}(2007)]%
        {woodruff2007evaluation}
\bibfield{author}{\bibinfo{person}{Susan~I Woodruff}, \bibinfo{person}{Terry~L
  Conway}, \bibinfo{person}{Christine~C Edwards}, \bibinfo{person}{Sean~P
  Elliott}, {and} \bibinfo{person}{Jim Crittenden}.}
  \bibinfo{year}{2007}\natexlab{}.
\newblock \showarticletitle{Evaluation of an Internet virtual world chat room
  for adolescent smoking cessation}.
\newblock \bibinfo{journal}{\emph{Addictive behaviors}} \bibinfo{volume}{32},
  \bibinfo{number}{9} (\bibinfo{year}{2007}), \bibinfo{pages}{1769--1786}.
\newblock


\end{thebibliography}
\appendix
\clearpage



\section{Help-Seeking Messages}
\label{appendix-messages}
\edit{These are the full texts for the four help-seeking messages described in section 3.2.1 labeled by the topic of the help-seeking message.}
\begin{enumerate}
    \item \textbf{Relationships}: "Hey, I'm feeling pretty emotional right now. So, get this, my ex invited me over to his place and things got weird. He started touching me and kept asking for sex, even though he has a girlfriend. I don't know why he does this. It's messed up, but I'm ashamed to admit that I'm kind of falling for it. We were on the phone just now and he fell asleep. Ugh, I can't believe this is happening."
    \item \textbf{Suicidal Thoughts}: "I don't know what to do. I'm having a panic attack and these thoughts won't stop. I feel so overwhelmed and trapped by my own mind. I don't feel like I'm worth anything and I don't see a way out. I'm scared and I don't know how to cope with these feelings. It's like I'm drowning, and I can't breathe. I don't want to live like this"
    \item \textbf{Self-Expression}: "I don't know what's wrong with me, I really struggle with communicating because of my anxiety, I don’t want to be judged and I always get so nervous when I have to share my thoughts and feelings with others. I'm afraid of not being taken seriously. I can’t seem to shake it off, I just wish I could get out of my head and stop overthinking everything."
    \item \textbf{Physical Health}: "I’m feeling drained both physically and emotionally... like, every morning, I wake up feeling tired and unmotivated. The thought of getting out of bed and facing the day seems like an impossible task. The other day, I went to counseling hoping it would help me feel better, but I feel the same. I still feel overwhelmed and stressed. All I want to do is to take some time to rest and recharge my energy."
\end{enumerate}

\section{Responses}
\label{appendix-responses}
\edit{These are the full texts for the 16 help-seeking message x response combinations described in section 3.2.1. Their labels are structured as follows: [topic of the help-seeking message] - [source of the response]}.
\begin{enumerate}
    \item \edit{\textbf{Relationships - Peer}}: “Tell them how you are feeling, and if they don’t understand and carry-on block them, even your ex-friends, you don’t need that in your life, be true to who you are, so do whatever feels right.”
    \item \edit{\textbf{Relationships - Adult Mentor}}: “I am so sorry you are feeling frustrated and hurt and you have every right to feel that way. I can tell you have been really struggling with your relationships and it's natural to feel attached to someone because we are all humans, and we all seek connection and the feeling of being understood and cared for. However, the relationship can become unhealthy if someone oversteps your boundaries and intentionally hurts you or others. Even if someone disrespects you and makes you feel invalidated, your feelings and thoughts still matter. It is really inspiring to see someone so resilient to share these personal thoughts and seek support, and I am here to listen.”
    \item \edit{\textbf{Relationships - Therapist}}: “Thank you for sharing with me, what I can tell is a complicated and difficult situation for you. Just to make sure I understand everything correctly, it sounds like you have mixed feelings about what is currently going on with your ex. On the one hand, I heard you say that you feel some shame or guilt for engaging in a recent sexual situation with him because he currently has a girlfriend. But I also heard you mention that you still have strong feelings for him. And that these recent experiences, including the phone call, have left you feeling quite emotional. Assuming I have that correct, I was hoping you could tell me more about the mixed emotions you have been feeling and then we can start to think through some next steps or strategies for you.”
    \item \edit{\textbf{Relationships - AI}}: "Hey, I'm sorry to hear you're going through such a tough situation. It can be really confusing when our emotions are at odds with what we know is right. Remember, your well-being and respect are important. Your ex's behavior is unacceptable, especially since he has a girlfriend. You deserve to be with someone who treats you with care and respect. It might be helpful to set clear boundaries and take some time to reflect on your own feelings. Don't hesitate to reach out if you need someone to talk to. You're stronger than you think, and you'll find your way through this. Take care.”
    \item \edit{\textbf{Suicidal Thoughts - Peer}}: “I understand how you feel and I’m proud of you for being alive and breathing despite your struggles, know that I am always here for you. You’re def not alone in your feelings...”
    \item \edit{\textbf{Suicidal Thoughts - Adult Mentor}}: “You deserve to feel safe, and I care about your safety. Could you tell me more about what happened to cause you to feel this way? I am here for you, and it seems like you want to escape a situation that is making you feel a lot of negative emotions like pain and stress. It can really feel frustrating when you are doubting your self worth and when you feel your struggles are permanent. You are worthy of love, care, dreams, hopes, warmth and so much more and you continue to inspire and impact people and things everyday perhaps without being aware of your influence as well.”
    \item \edit{\textbf{Suicidal Thoughts - Therapist}}: “It sounds like you are really struggling. I am glad that you reached out to me so I can offer help. There are some questions I need to ask you to learn more about how you are doing and to make sure you are safe. Are you having any thoughts of suicide or self-harm?”
    \item \edit{\textbf{Suicidal Thoughts - AI}}: “I'm really sorry to hear that you're feeling this way, but I can't provide the help that you need. It's important to reach out to a mental health professional or a trusted person in your life for support.”
    \item \edit{\textbf{Self-Expression - Peer}}: “Yeah, I think we all have those moments, you shouldn’t let it hold you back tho, be brave and put yourself out there.”
    \item \edit{\textbf{Self-Expression - Adult Mentor}}: “I get the feeling that you are anxious and frustrated with being able to fully express yourself without having to worry about how others think of you. If I understand you right, you feel hurt when people can’t fully listen to what you have to say, and it can even feel worse when you start to overthink self-worth. Your true authentic self has always been good enough and it’s okay if someone doesn’t like you, but it’s not okay if someone hurts you intentionally. You are not alone.”
    \item \edit{\textbf{Self-Expression - Therapist}}: “Thank you for sharing with me. What I can tell you has been a big challenge for you in terms of effectively communicating with others. I have worked with other clients in the past who also expressed challenges with communicating with others – including feeling anxious, judged, and nervous. It can be hard when you get stuck in your own thoughts – especially when many of your thoughts feel unhelpful or negative. There are strategies that you and I can work on together to address and change some of those negative thoughts so that you can feel more comfortable and confident when communicating with other people.”
    \item \edit{\textbf{Self-Expression - AI}}: "Hey, I'm sorry to hear that you're struggling with expressing yourself. I want you to know that your thoughts and feelings are important, and you deserve to be heard. It's common to feel anxious about being judged, but remember that your voice matters. Take small steps to overcome this, like sharing with a trusted friend or writing down your thoughts. Practice self-compassion and remember that you're not alone in feeling this way. You have valuable insights to share, and people who care will listen. Keep working on it, and don't be too hard on yourself. You're capable of breaking free from overthinking. Take care!”
    \item \edit{\textbf{Physical Health - Peer}}: “These things happen! Do you want to talk about it, How can I help you?”
    \item \edit{\textbf{Physical Health - Adult Mentor}}: “That must be so hard, having to balance everything related to schoolwork and life and feeling like you are carrying so much responsibility. There’s no wonder that waking up every day can be so challenging because there is so much burden to carry with everything going on. However, I can see you are trying to make progress and sometimes just getting out of bed is a big accomplishment. You are doing great, and it makes sense you want to spend more time resting since you exerted so much time in obligations that drained your energy.”
    \item \edit{\textbf{Physical Health - Therapist}}: “I am sorry you have been feeling so drained lately. When someone feels that way, it can be hard to feel like you have any energy, as you noted. I am pleased to hear you were able to make and attend a counseling appointment. I know it can feel disappointing to attend a session and not feel better after the session. But I do want to encourage you to keep trying counseling as it can sometimes take multiple sessions before people feel better. Hopefully, the counseling can help you reduce your stress and feelings of being overwhelmed.”
    \item \edit{\textbf{Physical Health - AI}}: "Hey there, I'm sorry to hear that you're feeling so drained and overwhelmed. It's completely normal to experience burnout, especially with schoolwork and life responsibilities. Remember, it's okay to take a step back and prioritize self-care. Resting and recharging your energy is important for your well-being. Counseling can take time to show its full effects, so don't be discouraged if you haven't seen immediate changes. Keep exploring different strategies and techniques that work best for you. Reach out to supportive friends or family members who can offer a listening ear. You're not alone in this, and you'll find your way through. Take care and give yourself the rest you need.” 
\end{enumerate}

\section{Context for ChatGPT Prompt}
\label{appendix-context}

\begin{enumerate}
    \item \textbf{Relationships}: Your friend is in a complicated relationship. They’ve sent you a message saying they're with a guy who already has a girlfriend, and they have feelings for him. They have no clue how to deal with this situation and how things will go. There may be a conflict between their mind and their heart: they know they’re doing the wrong thing, but they are following what they feel.
    \item \textbf{Suicidal Thoughts}: Your friend has a panic attack and is about to commit suicide. They’ve contacted you saying they’re in crisis and have no idea what to do to get out of it. They feel bad about themselves and think that they don’t deserve to live.
    \item \textbf{Self-Expression}: Your friend has a problem with expressing their real thoughts and feelings. They told you over text about their anxiety and are always scared of not being listened to or scared of being judged. Their overthinking keeps them stuck in thoughts and prevents them from doing a lot of things in life. Something holds them back every time they want to express themselves.
    \item \textbf{Physical Health}: Your friend feels drained and burnt out with schoolwork and life. They don’t have much energy to get up every morning and start their day. They are sharing their experience about counseling with you. They feel overwhelmed and just want to spend some time resting.
\end{enumerate}

\section{Data Distribution}

\begin{table}[h!]
\centering
\begin{tabular}{|c|c|c|c|c|c|}
\hline
\textbf{Source} & \textbf{Attribute} & \textbf{Mean} & \textbf{SD} & \textbf{Min} & \textbf{Max} \\ \hline
Peer            & like               & 3.03          & 1.14        & 1.00         & 5.00         \\ \hline
Peer            & problematic        & 2.32          & 1.24        & 1.00         & 5.00         \\ \hline
Peer            & would send         & 3.00          & 1.18        & 1.00         & 5.00         \\ \hline
Peer            & inappropriate      & 2.30          & 1.24        & 1.00         & 5.00         \\ \hline
Peer            & helpful            & 3.02          & 1.15        & 1.00         & 5.00         \\ \hline
Adult Mentor    & like               & 3.51          & 0.97        & 1.00         & 5.00         \\ \hline
Adult Mentor    & problematic        & 2.07          & 1.17        & 1.00         & 5.00         \\ \hline
Adult Mentor    & would send         & 3.48          & 0.98        & 1.00         & 5.00         \\ \hline
Adult Mentor    & inappropriate      & 2.05          & 1.19        & 1.00         & 5.00         \\ \hline
Adult Mentor    & helpful            & 3.50          & 0.99        & 1.00         & 5.00         \\ \hline
Therapist       & like               & 3.29          & 1.06        & 1.00         & 5.00         \\ \hline
Therapist       & problematic        & 2.19          & 1.22        & 1.00         & 5.00         \\ \hline
Therapist       & would send         & 3.25          & 1.12        & 1.00         & 5.00         \\ \hline
Therapist       & inappropriate      & 2.18          & 1.23        & 1.00         & 5.00         \\ \hline
Therapist       & helpful            & 3.29          & 1.07        & 1.00         & 5.00         \\ \hline
AI              & like               & 3.46          & 1.10        & 1.00         & 5.00         \\ \hline
AI              & problematic        & 2.13          & 1.22        & 1.00         & 5.00         \\ \hline
AI              & would send         & 3.42          & 1.12        & 1.00         & 5.00         \\ \hline
AI              & inappropriate      & 2.12          & 1.24        & 1.00         & 5.00         \\ \hline
AI              & helpful            & 3.49          & 1.05        & 1.00         & 5.00         \\ \hline
\end{tabular}
\caption{Data distribution for all scenarios combined (RQ1). For all variables, N = 2488 (622 participants x 4 scenarios).}
\end{table}

\begin{table*}[h!]
\centering
\begin{tabular}{|p{0.75cm}|ccc|ccc|ccc|ccc|}
\hline
                                       & \multicolumn{3}{c|}{\textbf{Relationships}}                                                & \multicolumn{3}{c|}{\textbf{Suicidal Thoughts}}                                            & \multicolumn{3}{c|}{\textbf{Self-Expression}}                                              & \multicolumn{3}{c|}{\textbf{Physical Health}}                                              \\ \hline
\textit{Source}                        & \multicolumn{1}{c|}{\textit{Attribute}} & \multicolumn{1}{c|}{\textit{Mean}} & \textit{SD} & \multicolumn{1}{c|}{\textit{Attribute}} & \multicolumn{1}{c|}{\textit{Mean}} & \textit{SD} & \multicolumn{1}{c|}{\textit{Attribute}} & \multicolumn{1}{c|}{\textit{Mean}} & \textit{SD} & \multicolumn{1}{c|}{\textit{Attribute}} & \multicolumn{1}{c|}{\textit{Mean}} & \textit{SD} \\ \hline
\multirow{5}{*}{\rotatebox[origin=c]{90}{\textbf{Peer}}}         & \multicolumn{1}{c|}{like}               & \multicolumn{1}{c|}{3.00}          & 1.15        & \multicolumn{1}{c|}{like}               & \multicolumn{1}{c|}{3.22}          & 1.08        & \multicolumn{1}{c|}{like}               & \multicolumn{1}{c|}{3.08}          & 1.09        & \multicolumn{1}{c|}{like}               & \multicolumn{1}{c|}{2.83}          & 1.19        \\ \cline{2-13} 
                                       & \multicolumn{1}{c|}{problematic}        & \multicolumn{1}{c|}{2.41}          & 1.24        & \multicolumn{1}{c|}{problematic}        & \multicolumn{1}{c|}{2.18}          & 1.24        & \multicolumn{1}{c|}{problematic}        & \multicolumn{1}{c|}{2.25}          & 1.19        & \multicolumn{1}{c|}{problematic}        & \multicolumn{1}{c|}{2.46}          & 1.28        \\ \cline{2-13} 
                                       & \multicolumn{1}{c|}{would send}         & \multicolumn{1}{c|}{2.98}          & 1.18        & \multicolumn{1}{c|}{would send}         & \multicolumn{1}{c|}{3.19}          & 1.12        & \multicolumn{1}{c|}{would send}         & \multicolumn{1}{c|}{3.05}          & 1.15        & \multicolumn{1}{c|}{would send}         & \multicolumn{1}{c|}{2.78}          & 1.23        \\ \cline{2-13} 
                                       & \multicolumn{1}{c|}{inappropriate}      & \multicolumn{1}{c|}{2.38}          & 1.24        & \multicolumn{1}{c|}{inappropriate}      & \multicolumn{1}{c|}{2.17}          & 1.22        & \multicolumn{1}{c|}{inappropriate}      & \multicolumn{1}{c|}{2.26}          & 1.23        & \multicolumn{1}{c|}{inappropriate}      & \multicolumn{1}{c|}{2.40}          & 1.27        \\ \cline{2-13} 
                                       & \multicolumn{1}{c|}{helpful}            & \multicolumn{1}{c|}{3.07}          & 1.13        & \multicolumn{1}{c|}{helpful}            & \multicolumn{1}{c|}{3.18}          & 1.11        & \multicolumn{1}{c|}{helpful}            & \multicolumn{1}{c|}{3.12}          & 1.09        & \multicolumn{1}{c|}{helpful}            & \multicolumn{1}{c|}{2.73}          & 1.22        \\ \hline
\multirow{5}{*}{\rotatebox[origin=c]{90}{\textbf{\begin{tabular}[c]{@{}c@{}}Adult\\ Mentor\end{tabular}}}} & \multicolumn{1}{c|}{like}               & \multicolumn{1}{c|}{3.48}          & 1.01        & \multicolumn{1}{c|}{like}               & \multicolumn{1}{c|}{3.59}          & 0.95        & \multicolumn{1}{c|}{like}               & \multicolumn{1}{c|}{3.51}          & 0.94        & \multicolumn{1}{c|}{like}               & \multicolumn{1}{c|}{3.48}          & 0.99        \\ \cline{2-13} 
                                       & \multicolumn{1}{c|}{problematic}        & \multicolumn{1}{c|}{2.04}          & 1.13        & \multicolumn{1}{c|}{problematic}        & \multicolumn{1}{c|}{2.04}          & 1.18        & \multicolumn{1}{c|}{problematic}        & \multicolumn{1}{c|}{2.10}          & 1.16        & \multicolumn{1}{c|}{problematic}        & \multicolumn{1}{c|}{2.10}          & 1.22        \\ \cline{2-13} 
                                       & \multicolumn{1}{c|}{would send}         & \multicolumn{1}{c|}{3.43}          & 1.02        & \multicolumn{1}{c|}{would send}         & \multicolumn{1}{c|}{3.56}          & 0.95        & \multicolumn{1}{c|}{would send}         & \multicolumn{1}{c|}{3.48}          & 0.97        & \multicolumn{1}{c|}{would send}         & \multicolumn{1}{c|}{3.44}          & 0.99        \\ \cline{2-13} 
                                       & \multicolumn{1}{c|}{inappropriate}      & \multicolumn{1}{c|}{2.04}          & 1.21        & \multicolumn{1}{c|}{inappropriate}      & \multicolumn{1}{c|}{2.00}          & 1.21        & \multicolumn{1}{c|}{inappropriate}      & \multicolumn{1}{c|}{2.08}          & 1.18        & \multicolumn{1}{c|}{inappropriate}      & \multicolumn{1}{c|}{2.06}          & 1.16        \\ \cline{2-13} 
                                       & \multicolumn{1}{c|}{helpful}            & \multicolumn{1}{c|}{3.47}          & 0.99        & \multicolumn{1}{c|}{helpful}            & \multicolumn{1}{c|}{3.61}          & 0.94        & \multicolumn{1}{c|}{helpful}            & \multicolumn{1}{c|}{3.51}          & 0.99        & \multicolumn{1}{c|}{helpful}            & \multicolumn{1}{c|}{3.43}          & 1.01        \\ \hline
\multirow{5}{*}{\rotatebox[origin=c]{90}{\textbf{Therapist}}}    & \multicolumn{1}{c|}{like}               & \multicolumn{1}{c|}{3.29}          & 1.09        & \multicolumn{1}{c|}{like}               & \multicolumn{1}{c|}{2.99}          & 1.12        & \multicolumn{1}{c|}{like}               & \multicolumn{1}{c|}{3.46}          & 0.97        & \multicolumn{1}{c|}{like}               & \multicolumn{1}{c|}{3.43}          & 0.97        \\ \cline{2-13} 
                                       & \multicolumn{1}{c|}{problematic}        & \multicolumn{1}{c|}{2.23}          & 1.24        & \multicolumn{1}{c|}{problematic}        & \multicolumn{1}{c|}{2.42}          & 1.27        & \multicolumn{1}{c|}{problematic}        & \multicolumn{1}{c|}{2.05}          & 1.15        & \multicolumn{1}{c|}{problematic}        & \multicolumn{1}{c|}{2.06}          & 1.19        \\ \cline{2-13} 
                                       & \multicolumn{1}{c|}{would send}         & \multicolumn{1}{c|}{3.24}          & 1.15        & \multicolumn{1}{c|}{would send}         & \multicolumn{1}{c|}{2.94}          & 1.20        & \multicolumn{1}{c|}{would send}         & \multicolumn{1}{c|}{3.42}          & 1.05        & \multicolumn{1}{c|}{would send}         & \multicolumn{1}{c|}{3.38}          & 1.01        \\ \cline{2-13} 
                                       & \multicolumn{1}{c|}{inappropriate}      & \multicolumn{1}{c|}{2.20}          & 1.24        & \multicolumn{1}{c|}{inappropriate}      & \multicolumn{1}{c|}{2.42}          & 1.30        & \multicolumn{1}{c|}{inappropriate}      & \multicolumn{1}{c|}{2.09}          & 1.19        & \multicolumn{1}{c|}{inappropriate}      & \multicolumn{1}{c|}{2.02}          & 1.16        \\ \cline{2-13} 
                                       & \multicolumn{1}{c|}{helpful}            & \multicolumn{1}{c|}{3.27}          & 1.07        & \multicolumn{1}{c|}{helpful}            & \multicolumn{1}{c|}{2.97}          & 1.15        & \multicolumn{1}{c|}{helpful}            & \multicolumn{1}{c|}{3.49}          & 0.98        & \multicolumn{1}{c|}{helpful}            & \multicolumn{1}{c|}{3.45}          & 0.99        \\ \hline
\multirow{5}{*}{\rotatebox[origin=c]{90}{\textbf{AI}}}           & \multicolumn{1}{c|}{like}               & \multicolumn{1}{c|}{3.63}          & 1.00        & \multicolumn{1}{c|}{like}               & \multicolumn{1}{c|}{2.82}          & 1.20        & \multicolumn{1}{c|}{like}               & \multicolumn{1}{c|}{3.70}          & 0.98        & \multicolumn{1}{c|}{like}               & \multicolumn{1}{c|}{3.71}          & 0.96        \\ \cline{2-13} 
                                       & \multicolumn{1}{c|}{problematic}        & \multicolumn{1}{c|}{2.01}          & 1.16        & \multicolumn{1}{c|}{problematic}        & \multicolumn{1}{c|}{2.53}          & 1.26        & \multicolumn{1}{c|}{problematic}        & \multicolumn{1}{c|}{2.01}          & 1.22        & \multicolumn{1}{c|}{problematic}        & \multicolumn{1}{c|}{1.95}          & 1.16        \\ \cline{2-13} 
                                       & \multicolumn{1}{c|}{would send}         & \multicolumn{1}{c|}{3.65}          & 0.96        & \multicolumn{1}{c|}{would send}         & \multicolumn{1}{c|}{2.71}          & 1.21        & \multicolumn{1}{c|}{would send}         & \multicolumn{1}{c|}{3.67}          & 1.00        & \multicolumn{1}{c|}{would send}         & \multicolumn{1}{c|}{3.64}          & 0.99        \\ \cline{2-13} 
                                       & \multicolumn{1}{c|}{inappropriate}      & \multicolumn{1}{c|}{2.01}          & 1.20        & \multicolumn{1}{c|}{inappropriate}      & \multicolumn{1}{c|}{2.54}          & 1.27        & \multicolumn{1}{c|}{inappropriate}      & \multicolumn{1}{c|}{1.99}          & 1.18        & \multicolumn{1}{c|}{inappropriate}      & \multicolumn{1}{c|}{1.95}          & 1.21        \\ \cline{2-13} 
                                       & \multicolumn{1}{c|}{helpful}            & \multicolumn{1}{c|}{3.63}          & 0.95        & \multicolumn{1}{c|}{helpful}            & \multicolumn{1}{c|}{2.93}          & 1.12        & \multicolumn{1}{c|}{helpful}            & \multicolumn{1}{c|}{3.75}          & 0.94        & \multicolumn{1}{c|}{helpful}            & \multicolumn{1}{c|}{3.67}          & 0.99        \\ \hline
\end{tabular}
\caption{Data distribution for individual scenarios (RQ2). For all variables, min=1.00, max=5.00, and N=622.}
\label{tab:distribution-scenarios}
\end{table*}





\clearpage
\end{document}